\documentclass[reprint,superscriptaddress,nofootinbib,twocolumn,amsmath,amssymb,aps,prl]{revtex4-2}

\usepackage{graphicx}
\usepackage{dcolumn}
\usepackage{float}
\usepackage{bm}
\usepackage[colorlinks]{hyperref}
\usepackage[utf8]{inputenc}	
\usepackage{lmodern} 
\usepackage{siunitx}
\usepackage[version=4]{mhchem}
\usepackage{microtype}
\usepackage{verbatim}
\usepackage{lineno}
\usepackage[normalem]{ulem}

\newcommand{\bra}[1]{\langle#1|}
\newcommand{\ket}[1]{|#1\rangle}
\newcommand{\braket}[1]{\langle#1\rangle}

\usepackage{gensymb}
\usepackage{xcolor}
\usepackage{amsthm}
\usepackage{amsfonts}
\usepackage{epsfig}
\usepackage{cancel}

\begin{document}
\title{Measuring the Quantum Phase of Multiphoton Systems}

\author{Ian Baum}
\affiliation{Quantum Photonics Laboratory, Department of Physics \& Astronomy, Louisiana State University, Baton Rouge, LA 70803, USA}

\author{Riley B. Dawkins}
\email{rdawki1@lsu.edu}
\affiliation{Quantum Photonics Laboratory, Department of Physics \& Astronomy, Louisiana State University, Baton Rouge, LA 70803, USA}

\author{Jannatul Ferdous}
\affiliation{Quantum Photonics Laboratory, Department of Physics \& Astronomy, Louisiana State University, Baton Rouge, LA 70803, USA}

\author{Omar S. Maga\~na-Loaiza}
\affiliation{Quantum Photonics Laboratory, Department of Physics \& Astronomy, Louisiana State University, Baton Rouge, LA 70803, USA}

\date{\today}

\begin{abstract}
The quantum phase of a multiparticle system has remained a long-standing mystery in quantum physics. The absence of a phase operator that faithfully represents this physical quantity has prevented the formulation of a meaningful uncertainty relation with particle number. Here, we resolve this problem by measuring, for the first time, the quantum phase of multiparticle systems containing up to fifteen particles. We achieve this through a new quantum state tomography scheme that combines homodyne interferometry with photon-number-resolving detection. This approach enables us to reconstruct complementary photon-number and phase representations and establish the corresponding number–phase uncertainty relation. Our measurements further provide access to the mutually unbiased basis of a multiparticle system, a fundamental resource for multiparticle quantum information processing. Furthermore, we report the spatially resolved characterization of the quantum phase. These results provide an operational framework for measuring quantum phase in multiparticle systems, with implications for fundamental quantum physics and scalable multiparticle quantum technologies.

\end{abstract}

\maketitle

Phase is one of the most fundamental quantities in physics, underlying interference and our description of wave phenomena \cite{CarruthersRMP1968, SpechtNP2009, ChrapkiewiczNatPhoton2016}. In quantum mechanics, however, the physical meaning of phase becomes considerably more subtle \cite{PeggJMO1990, SusskindPRF1964}. While particle number is represented by a well-defined quantum observable, the quantum description of its conjugate variable, phase, has remained a long-standing problem \cite{LynchPR1995, MandelWolfRMP1965}. Despite decades of theoretical efforts, the absence of a phase operator that faithfully represents this physical quantity has prevented the direct measurement of multiparticle phase states and the formulation of a meaningful uncertainty relation with particle number \cite{MandelPR1964, NohPRL1991}. The quantum phase of a multiparticle system therefore remains an enduring question in quantum physics \cite{PeggJMO1990, SusskindPRF1964, LynchPR1995, MandelPR1964, NohPRL1991, MandelWolfRMP1965}.

This problem has acquired new significance with the emergence of many-body photonic systems and their use in multiparticle quantum technologies \cite{WalmsleyOpticaQ2023, MagañaLoaizaNQI2019, AmicoRMP2022, YouNature2026, YouNatCommun2021, ZhangAOP2024}. The large number of simultaneous interactions that these systems can host has enabled powerful schemes for imaging, sensing, and quantum information processing \cite{MostafaviAPR2025, HongAdvSci2026, FerdousACSP2024, StobinskaSciAdv2019, YouAPR2020, AspuruGuzikNatPhys2012}. Many of these protocols rely on access to complementary variables, which can be readily obtained through Fourier transformations for conventional pairs such as position and momentum, time and energy, or angular position and orbital angular momentum \cite{HowellPRL2004, LeachScience2010, XavierNQ2025, YouPhotonix2024, HuhNatPhoton2015, ZhangNatCommun2012, Erhard2020}. Such an approach remains inaccessible for particle number and phase \cite{HongAdvSci2026, StobinskaSciAdv2019}. Even leaving aside the demanding runtimes associated with implementing a discrete Fourier transform in the number-state basis, the lack of a physically meaningful quantum phase has prevented access to phase states in multiphoton systems \cite{StobinskaSciAdv2019,WalmsleyOpticaQ2023, CrespiNatCommun2016, AmicoRMP2022}. Because photon-number and phase states are related through a Fourier transformation, these states provide complementary, mutually unbiased bases for encoding and measuring multiparticle quantum information \cite{WoottersAnnPhys1989, AdamsonPRL2010}. Measuring these states would therefore establish the number--phase uncertainty relation and provide experimental access to mutually unbiased bases, opening new possibilities for multiparticle quantum metrology, sensing, and information processing \cite{ BhusalNPJ2022, MostafaviAPR2025, HongAdvSci2026, ZhangAOP2024, AspuruGuzikNatPhys2012, WalmsleyOpticaQ2023, AmicoRMP2022}.

Here, we report the first measurement of multiparticle quantum phase states and characterize multiphoton fields in the complementary variables of particle number and phase. To achieve this, we developed a new quantum state tomography scheme that combines homodyne interferometry with photon-number-resolving detection, enabling us to reconstruct phase states containing up to fifteen particles  \cite{VogelPRA1989, YouAPR2020}. These measurements establish the corresponding number--phase uncertainty relation and provide access to the mutually unbiased basis of multiparticle systems. Furthermore, we demonstrate the spatially resolved characterization of quantum phase. Together, these results provide experimental access to a fundamental quantity that has remained elusive throughout the history of quantum mechanics and open new possibilities for fundamental quantum physics and multiparticle quantum technologies.

\begin{figure*}
    \centering
\includegraphics[width=0.69\paperwidth]{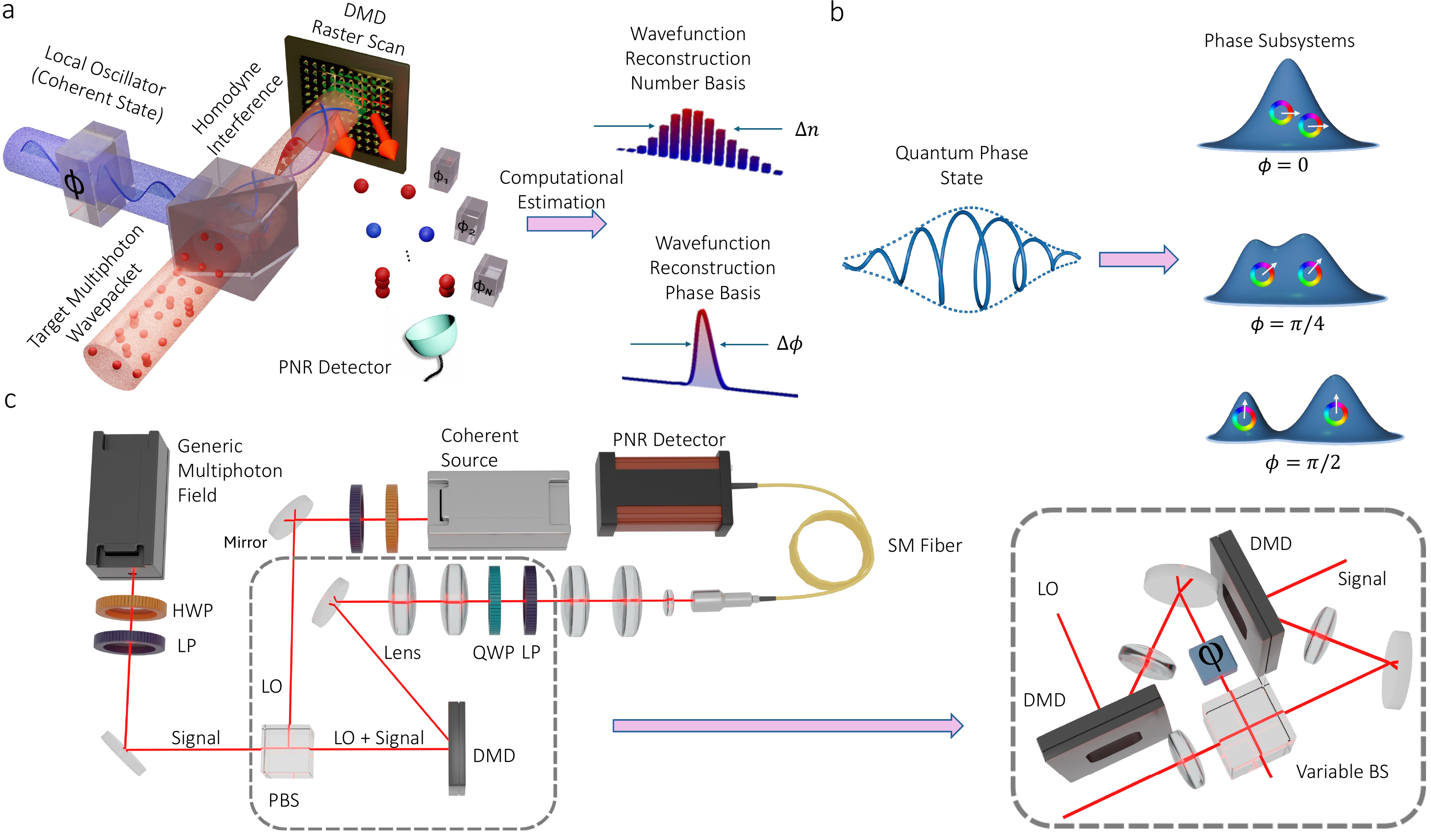}
    \caption{\textbf{Spatially-photon-number-resolved homodyne tomography}. As illustrated in \textbf{a}, we employ a new quantum state tomography scheme that combines homodyne interferometry with photon-number-resolving (PNR) detection to reconstruct the complementary photon-number and phase bases of an optical field. This information enables us to select both the constituent multiphoton (photon-number) subsystems that form an optical field and multiphoton phase states with specific phase values. As shown in \textbf{b}, this capability enables us to extract multiphoton phase states with well-defined phases from an arbitrary optical field. The experimental implementation is shown in \textbf{c}. The unknown signal and a coherent local oscillator (LO) are prepared at 532~nm in orthogonal polarizations and combined at a polarizing beam splitter (PBS). A digital micromirror device (DMD) performs a transverse spatial raster scan. The output passes through a variable zero-order quarter-wave plate (QWP) and a linear polarizer before being coupled into a single-mode fiber and measured with a PNR detector. The right schematic in \textbf{c} illustrates the equivalent homodyne picture: mixing the two polarizations with the QWP and linear polarizer controls the relative phase and amplitude of the signal and LO, analogous to interference between two single-mode fields at a variable beam splitter. Additional details of the experimental setup and measurement protocol are provided in the SI.}
    \label{fig:setup}
     \label{Figure1}
\end{figure*}


We demonstrate spatially resolved photon-number tomography using the experimental setup depicted in Fig.~\ref{Figure1}\textbf{a}. This setup enables the characterization of arbitrary multiphoton fields through a new homodyne-like approach that exploits photon-number-resolving (PNR) detection \cite{HongLPR2023, SmitheyPRL1993}. Here, an arbitrary, unknown signal state is combined with a coherent state, \textcolor{black}{$\ket{\beta}$}, serving as the local oscillator (LO) in the same spatial mode. To reconstruct the signal state, we vary both the relative phase and amplitude of the LO and signal using linear polarization optics parameterized by the angles $\vec{\theta}$. For each setting, the corresponding photon-number positive operator-valued measure (POVM) elements are obtained by tracing over the LO; a detailed derivation is provided in section 1 of the Supplementary Information (SI). To avoid instabilities associated with linear inversion, we reconstruct the target quantum state $\hat{\rho}$ by maximizing the likelihood associated with the measured photon-number, subject to the density matrix being positive semidefinite and having unit trace \cite{lvovsky2004iterative}.

\begin{equation} \label{likelihood}
    \rho_{\mathrm{ML}}
    =
    \underset{\rho \geq 0,\,
    \operatorname{Tr}\{\rho\}=1}{\arg\max}
    \sum_{\vec{\theta},n}
    N_n(\vec{\theta})
    \ln
    [
    \operatorname{Tr}
    \{
    \hat{\rho} \hat{\Gamma}_n(\vec{\theta})
    \}
    ].
\end{equation}
To perform this optimization, we employ the iterative diluted-$R\hat{\rho}R$ algorithm to reconstruct the density matrix independently at each transverse spatial position \cite{lvovsky2004iterative}. At each pixel, the algorithm progressively updates the density matrix toward higher-likelihood physical states until convergence is reached. Further details of the algorithm is provided in section 1 of the SI. 

Our measurement scheme for characterizing arbitrary multiphoton fields provides a direct route to exploring the number--phase problem, which can be formulated through a fundamental question: what physical quantity represents quantum phase? The classical phase of an electromagnetic field is well defined, but it is a classical property associated with the field amplitude rather than the quantum state itself \cite{NohPRL1991, NohPRA1992b}. Here, we instead seek a quantum phase encoded in the statistical properties of the optical field and capable of characterizing multiparticle phase states. This requires identifying a quantum observable whose measurement outcomes can be directly associated with phase. Constructing such an observable is fundamentally challenging because quantum observables are conventionally represented by self-adjoint operators with real-valued\textcolor{black}{, Euclidean} spectra, whereas phase is intrinsically periodic \textcolor{black}{because} values differing by integer multiples of $2\pi$ represent the same physical outcome \cite{NohPRL1991, NohPRA1992b}. A physically meaningful phase observable must therefore preserve this periodicity while remaining compatible with the mathematical structure of quantum mechanics. This motivates an operator whose eigenvalues are the complex phases $e^{i\phi}$, establishing a direct correspondence between each eigenvalue and a phase angle $\phi$. For a single-mode field, this operator \textcolor{black}{turns out to be given by} $\hat{E}=\sum_{n=0}^{\infty}\ket{n}\bra{n+1}$.

\begin{figure*}
    \centering
\includegraphics[width=0.76\paperwidth]{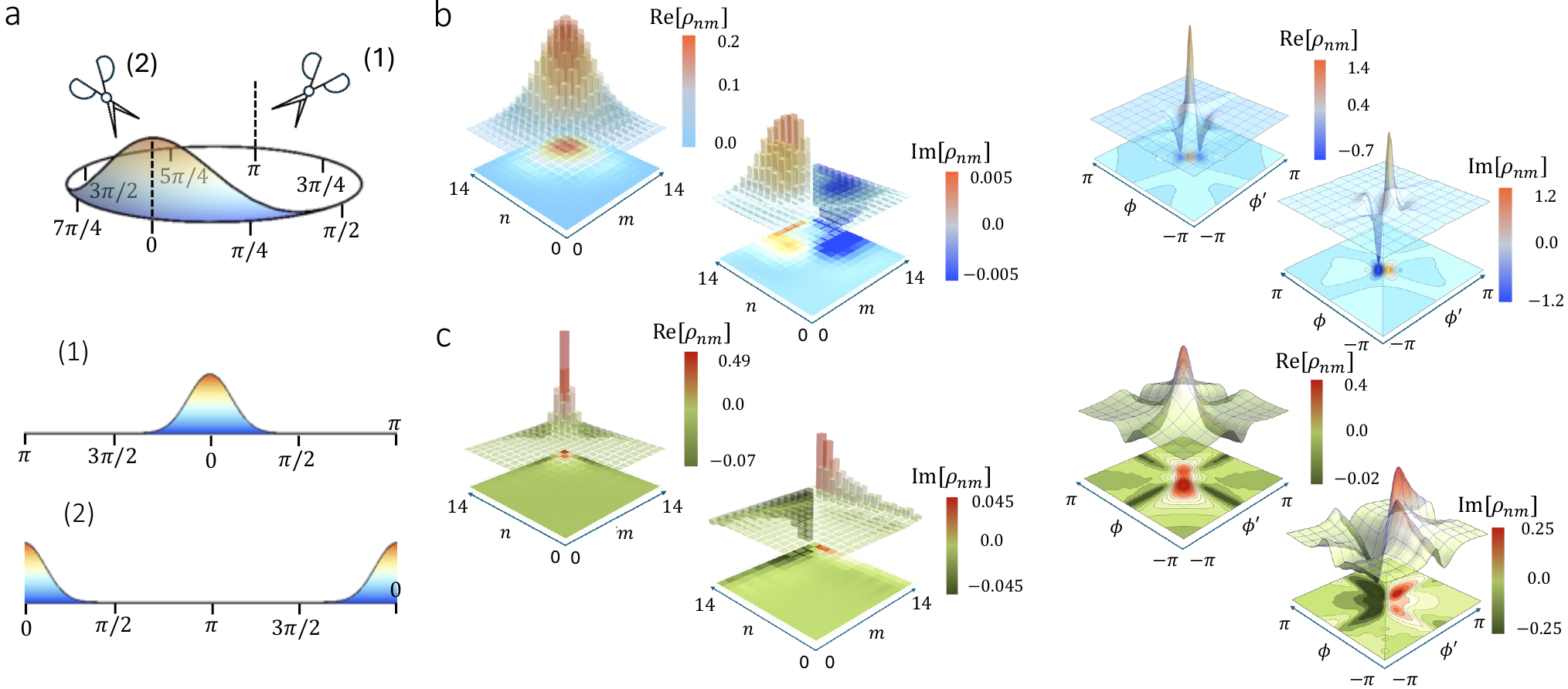}
    \caption{\textbf{
Quantum state tomography of multiphoton fields in the number and phase bases.}     The periodic nature of quantum phase prevents its direct characterization through a linear average of phase values. As illustrated in \textbf{a}, because phase is defined on a circle, the inferred mean can depend strongly on the arbitrary position of the phase reference, here shown for two different choices of the $2\pi$ boundary. The trigonometric observables $\hat{C}$ and $\hat{S}$ circumvent this ambiguity by providing a representation that preserves the periodicity of phase. Using these observables, our tomography enables the reconstruction of multiphoton fields in the complementary photon-number and phase bases. As shown in \textbf{b}, a coherent multiphoton field extends over multiple photon-number states while exhibiting a phase distribution strongly localized around a well-defined phase, with its narrow width reflecting a comparatively small phase uncertainty. In contrast, the displaced thermal multiphoton field shown in \textbf{c} exhibits enhanced photon-number fluctuations accompanied by a broader phase distribution and, consequently, a larger phase uncertainty, while remaining centered around a well-defined phase. These measurements illustrate how the widths of the complementary photon-number and phase distributions are governed by the statistical properties of multiphoton fields.}
    \label{fig:2}
\end{figure*}

Although $\hat{E}$ captures the periodic nature of phase, it is not Hermitian and therefore does not represent a conventional quantum observable \cite{MandelPR1964}. We therefore construct the Hermitian trigonometric phase observables $\hat{C}=\frac{1}{2}(\hat{E}+\hat{E}^\dagger)$ and $\hat{S}=\frac{1}{2i}(\hat{E}-\hat{E}^\dagger)$, which represent the cosine and sine components of the quantum phase, respectively. Further details of their algebra are provided in the SI.

Together, $\hat{C}$ and $\hat{S}$ provide an operational representation of quantum phase, allowing both its value and uncertainty to be determined from measurable quantities. We obtain the phase uncertainty by applying error propagation to these two observables. This construction preserves the intrinsic periodicity of phase and leads directly to the photon-number--phase uncertainty relation

\begin{equation}\label{Eq2}
\Delta n \Delta \phi \geq \sqrt{\frac{\frac{1}{4}\braket{\hat{S}}^4 + \frac{1}{4}\braket{\hat{C}}^4 - 2 [\Delta n]^2\braket{\hat{C}}\braket{\hat{S}}\mathrm{Cov}(\hat{C},\hat{S})}{(\braket{\hat{C}}^2+\braket{\hat{S}}^2)^2}}.
\end{equation}

Importantly, the lower bound is state dependent and is governed by the quantity $\mathcal{P}(\hat{\rho})=\langle\hat{C}\rangle\langle\hat{S}\rangle\mathrm{Cov}(\hat{C},\hat{S})$, bounded by $-\Delta C,\Delta S,|\langle\hat{C}\rangle\langle\hat{S}\rangle|\leq\mathcal{P}(\hat{\rho})\leq\Delta C,\Delta S,|\langle\hat{C}\rangle\langle\hat{S}\rangle|$. At \textcolor{black}{its minimum value}, $\mathcal{P}(\hat{\rho})=-\Delta C,\Delta S,|\langle\hat{C}\rangle\langle\hat{S}\rangle|$, the uncertainty bound reduces to $1/2$, recovering the familiar form of the position--momentum uncertainty relation. Other limiting cases, including states for which the lower bound vanishes, are discussed in section 2 of the SI.  Our tomography therefore provides simultaneous access to the photon-number and phase representations of an arbitrary multiphoton field, enabling us to reconstruct the phase distributions associated with its quantum statistical properties \cite{HongNatPhys2024,DawkinsOL2024}. By performing this reconstruction at each transverse position, we measure the spatial distribution of the number--phase uncertainty and reveal how local photon-number statistics determine the corresponding phase uncertainty across the optical field \cite{HongLPR2023, FerdousACSP2024, VarcoeNJP2004}.

Having established the number--phase uncertainty relation for a single-mode multiphoton field, we now extend this framework to many modes by considering a spatially resolved quantum field \cite{HongLPR2023, FerdousACSP2024}. Within quantum field theory, localized measurements of the electromagnetic field can be described by point detectors occupying a finite volume $V$, whose action is represented by the smeared annihilation operator \textcolor{black}{$\hat{a}_{s,V}(\mathbf{r})=\frac{1}{\sqrt{V}}\int_{B(\mathbf{r},V)}\frac{d^3r'}{(2\pi)^{3/2}}\hat{a}_s(\mathbf{r}')$}, where $B(\mathbf{r},V)$ denotes the volume $V$ centered at position $\mathbf{r}$ and $\hat{a}_s(\mathbf{r}')$ is the annihilation-operator density for polarization $s$ \cite{MandelPR1964}. This local description enables us to define the photon-number multiphoton subsystem operators $\hat{n}(N,M|\mathbf{r})$ and, crucially, the corresponding phase multiphoton subsystem operators $\hat{\Phi}(\phi|\mathbf{r})$,

\begin{equation}\label{Eq3}
    \begin{aligned}
        \hat{n}(N,M | \mathbf{r}) &= \textcolor{black}{:\left(\frac{\left(\hat{a}^\dagger_{s,V}(\mathbf{r})\right)^N\left(\hat{a}_{s,V}(\mathbf{r})\right)^M}{\sqrt{N!M!}}e^{-\hat{a}^\dagger_{s,V}(\mathbf{r})\hat{a}_{s,V}(\mathbf{r})}\right):,}\\
        \hat{\Phi}(\phi|\mathbf{r}) &= \sum_{n,m=0}^\infty e^{i\phi(n-m)}\hat{n}(n,m|\mathbf{r}).
    \end{aligned}
\end{equation}
The second expression defines a local multiphoton phase subsystem with phase $\phi$, as illustrated schematically in Fig.~\ref{Figure1}\textbf{b}, where multiphoton phase states with well-defined values of $\phi$ are extracted from the optical field. This definition establishes the physical quantity that we subsequently access experimentally through our tomographic scheme. Taking the expectation values of these operators over the quantum state $\hat{\rho}$ yields the local photon-number and phase subsystems, $\hat\rho(N|\mathbf{r})=\mathrm{Tr}[\hat{\rho}\hat{n}(N,N|\mathbf{r})]$ and $\hat\rho(\phi|\mathbf{r})=\mathrm{Tr}[\hat{\rho}\hat{\Phi}(\phi|\mathbf{r})]$, respectively. These distributions characterize the photon-number and phase states of the multiphoton subsystems constituting the optical field, giving the probabilities of observing photon number $N$ and phase $\phi$, respectively, at position $\mathbf{r}$. The ability to project multiphoton fields onto well-defined photon-number states has recently enabled the observation and manipulation of otherwise inaccessible multiparticle phenomena~\cite{YouNatCommun2021, BhusalNPJ2022,MostafaviAPR2025, HongNatPhys2024, YouNature2026, HongAdvSci2026}. Our framework extends this capability to the complementary phase basis, enabling projections onto multiphoton states with well-defined quantum phases. Given the central role of phase in quantum coherence and interference, such phase-selective measurements could expose multiparticle phenomena that remain hidden in the photon-number representation and provide new resources for coherent quantum technologies ~\cite{YouNatCommun2021, BhusalNPJ2022,MostafaviAPR2025, HongNatPhys2024, YouNature2026, HongAdvSci2026}. Together with the number--phase uncertainty relation established in \textcolor{black}{Eq.(\ref{Eq2})}, this framework provides the theoretical basis for experimentally reconstructing the spatially resolved photon-number and phase distributions, and their associated uncertainty, using our quantum state tomography \textcolor{black}{protocol}. Further details of the quantum-field-theoretical formulation and the numerical determination of these subsystems using the Glauber $P$ function are provided in section 3 of the SI \cite{DawkinsOL2024, GlauberPR1963, SudarshanPRL1963}.

We now describe the experimental implementation used to characterize the photon-number and phase states of optical fields with arbitrary quantum statistical properties \cite{HongLPR2023,YouNature2026}. Our photon-counting homodyne tomography combines homodyne interferometry with PNR detection, providing a unified framework for characterizing multiphoton fields with coherent, thermal, and superthermal statistics \cite{YouAPR2020, YouNature2026}. As shown in Fig.~\ref{Figure1}\textbf{c}, the signal field interferes with a coherent-state LO, whose polarization is transformed to independently control the relative phase and amplitude of the two fields \cite{VogelPRA1989, SmitheyPRL1993}. Varying these quantities generates an informationally complete set of phase-sensitive measurements for reconstructing the signal state. For each interferometric setting, a PNR detector records the complete photon-number statistics, providing the phase-sensitive photon-number distributions required to reconstruct the density matrix \cite{DawkinsOL2024}. To spatially resolve the quantum phase, the signal and LO are prepared in matching Gaussian spatial modes and raster scanned across the transverse plane using a digital micromirror device (DMD) \cite{BhusalNPJ2022, HongLPR2023}. The transformation introduced by the DMD is incorporated into the POVM elements used for the tomographic reconstruction, accounting for polarization-dependent effects during the measurement. The reconstructed density matrix then provides direct experimental access to the complementary photon-number and phase representations introduced above, enabling us to measure their spatial structure across multiphoton fields with distinct quantum statistical properties \cite{YouAPR2020, YouPhotonix2024, YouNature2026}. 

\begin{figure*}
    \centering
    \includegraphics[width=0.66\paperwidth]{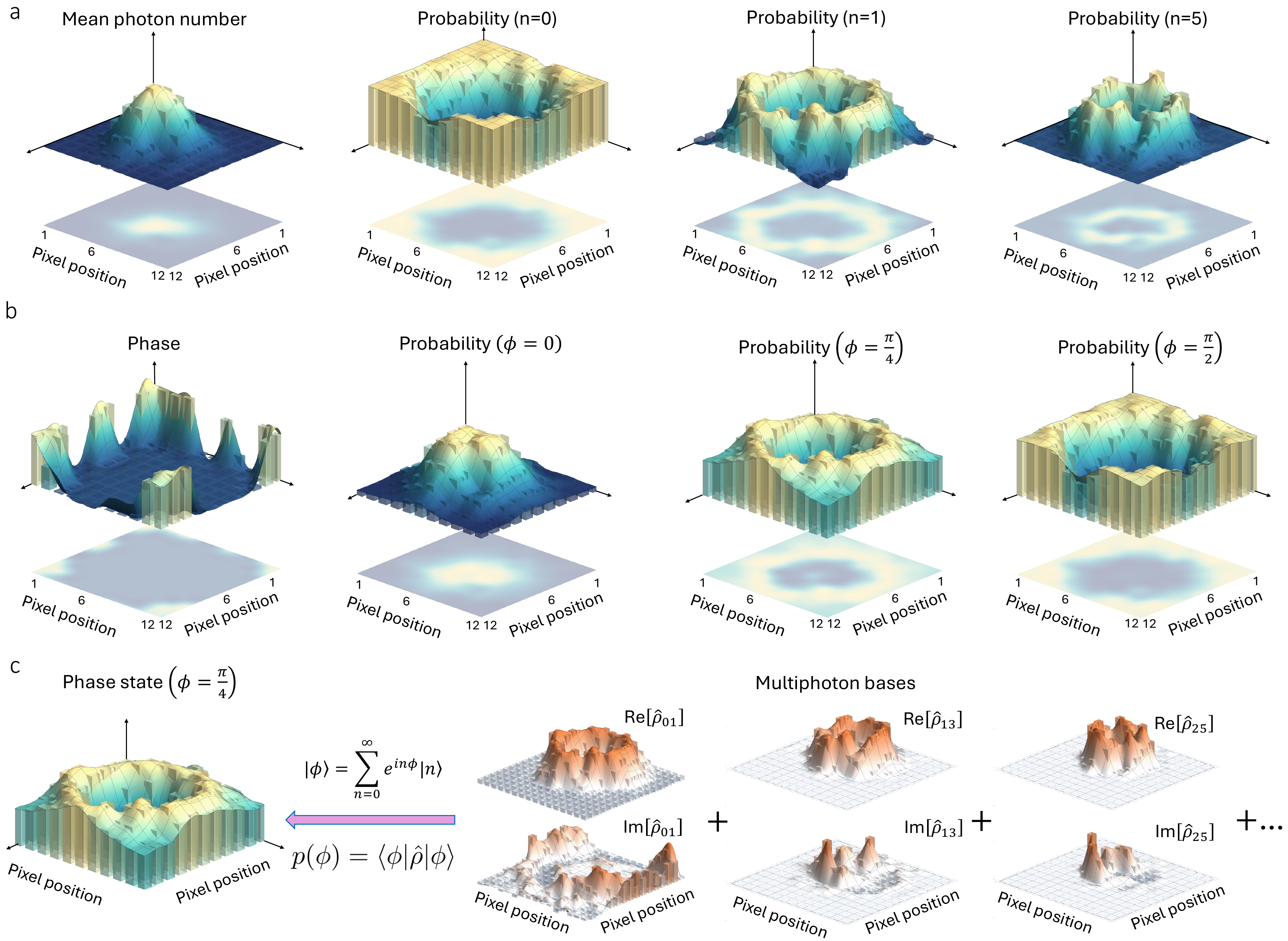}
    \caption{\textbf{Number--phase Fourier duality in spatially resolved multiphoton fields.} Our spatially resolved quantum state tomography reveals the complementary photon-number and phase structure of a multiphoton field. As shown in \textbf{a}, we demonstrate the possibility of distilling the constituent multiphoton subsystems of a classical light field from its mean photon-number (intensity) distribution. This capability is illustrated for $n=0$, 1, and 5, revealing distinct spatial distributions for each multiphoton subsystem. The complementary phase representation is shown in \textbf{b}, where the reconstructed quantum phase is resolved into multiphoton phase subsystems with well-defined values \textcolor{black}{$\phi=0$}, $\pi/4$, and $\pi/2$. These phase subsystems exhibit spatial distributions that evolve with \textcolor{black}{$\phi$}, revealing the phase structure encoded across the multiphoton field. The origin of this number--phase Fourier duality is illustrated in \textbf{c}. A multiphoton phase state with a well-defined phase, here \textcolor{black}{$\phi=\pi/4$}, arises from the coherent superposition of photon-number states, \textcolor{black}{$\ket{\phi}=\sum_{n=0}^{\infty}e^{in\phi}|n\rangle$}. Consequently, its spatial probability distribution, \textcolor{black}{$p(\phi)=\langle\phi|\hat{\rho}|\phi\rangle$}, is determined by the populations and coherences of the constituent multiphoton subsystems, represented by the density-matrix elements $\rho_{nm}$. Together, these measurements provide a spatially resolved realization of the Fourier duality between photon number and quantum phase, directly revealing how multiphoton coherences give rise to phase states across an optical field. This experimentally accessible Fourier relationship provides a direct route to mutually unbiased photon-number and phase bases for multiparticle quantum information processing.}
    \label{fig:3}
\end{figure*}

We first reconstruct the complementary photon-number and phase representations of multiphoton fields with distinct quantum statistical properties. As shown in Fig.~\ref{fig:2}\textbf{a}, the periodicity of phase makes a conventional linear average dependent on the arbitrary choice of phase reference, motivating the use of the trigonometric phase observables introduced above \cite{NohPRL1991,NohPRA1992, NohPRA1992b}. Using our photon-counting homodyne tomography \textcolor{black}{protocol}, we reconstruct both the real and imaginary components of the density matrix in the Fock and phase bases for a coherent multiphoton field containing up to 15 photons, as shown in Fig.~\ref{fig:2}\textbf{b}. The Fock-basis representation resolves the photon-number composition of the field, whereas the complementary phase representation reveals a well-localized distribution centered around a well-defined phase. We next investigate how this phase structure changes with the statistical properties of the multiphoton field. Figure~\ref{fig:2}\textbf{c} shows the corresponding reconstruction for a displaced thermal multiphoton field. In contrast to the coherent field, the displaced thermal field exhibits a substantially broader phase distribution, revealing an increased phase uncertainty associated with its thermal photon-number fluctuations \cite{HongLPR2023,YouNature2026}. This behavior is consistent with the limiting case of a purely thermal field, for which the absence of a preferred phase produces a uniform phase distribution with equal probability for all phase values. Together, these measurements demonstrate that our tomography provides simultaneous experimental access to photon-number statistics and the quantum phase information encoded in the coherences of multiphoton fields, while resolving how their phase distributions change with their quantum statistical properties.

We next investigate the spatial structure of the complementary photon-number and phase representations of a coherent multiphoton field \cite{GlauberPR1963, SudarshanPRL1963, DawkinsOL2024}. Photon-counting homodyne tomography is performed at each transverse position, enabling reconstruction of the local density matrix and, through Eq.~(\ref{Eq3}), extraction of the corresponding photon-number and phase subsystems. As shown in Fig.~\ref{fig:3}\textbf{a}, the mean photon number exhibits the Gaussian spatial profile of the coherent field, while its constituent photon-number subsystems display markedly different spatial structures for $n=0$, 1, and 5 \cite{DawkinsOL2024, HongNatPhys2024}. The vacuum subsystem is suppressed near the beam center, where the mean photon number is largest, whereas increasing photon number progressively localizes the corresponding multiphoton subsystem toward the center of the field. The complementary phase representation is shown in Fig.~\ref{fig:3}\textbf{b}. While the average phase remains approximately uniform across the field, post-selection of well-defined phases \textcolor{black}{$\phi=0$}, $\pi/4$, and $\pi/2$ reveals strongly phase-dependent spatial distributions. The probability of the coherent-state phase, \textcolor{black}{$\phi=0$}, is localized near the beam center, whereas increasing phase mismatch progressively suppresses this central contribution. These complementary spatial distributions reveal the reciprocal behavior of photon-number and phase fluctuations across the multiphoton field and originate from their underlying Fourier relationship \cite{PeggJMO1990, NohPRL1991}. As illustrated in Fig.~\ref{fig:3}\textbf{c} for \textcolor{black}{$\phi=\pi/4$}, a phase state \textcolor{black}{$\ket{\phi}=\sum_{n=0}^{\infty}e^{in\phi}|n\rangle$} is formed from a coherent superposition of photon-number states, such that its spatial probability \textcolor{black}{$p(\phi)=\langle\phi|\hat{\rho}|\phi\rangle$} contains contributions from both the populations and coherences $\rho_{nm}$ of the multiphoton field. To our knowledge, this constitutes the first experimental reconstruction of a spatially resolved multiphoton phase state from its constituent photon-number populations and coherences, providing a direct visualization of the Fourier relationship between particle number and quantum phase. Our measurements therefore provide a spatially resolved realization of number--phase Fourier duality. Specifically, the photon-number subsystems resolve the constituent multiphoton components of the field, while their coherent Fourier superposition gives rise to spatially structured phase subsystems with well-defined quantum phases \cite{CrespiNatCommun2016, StobinskaSciAdv2019}. This experimentally accessible Fourier relationship further provides a direct route to mutually unbiased photon-number and phase bases \cite{WoottersAnnPhys1989, Wiesniak2011}, enabling complementary multiphoton states to be prepared and measured without physically implementing a discrete Fourier transformation in the photon-number basis \cite{PeggJMO1990}.

\begin{figure}[t]
    \centering
    \includegraphics[width=\columnwidth]{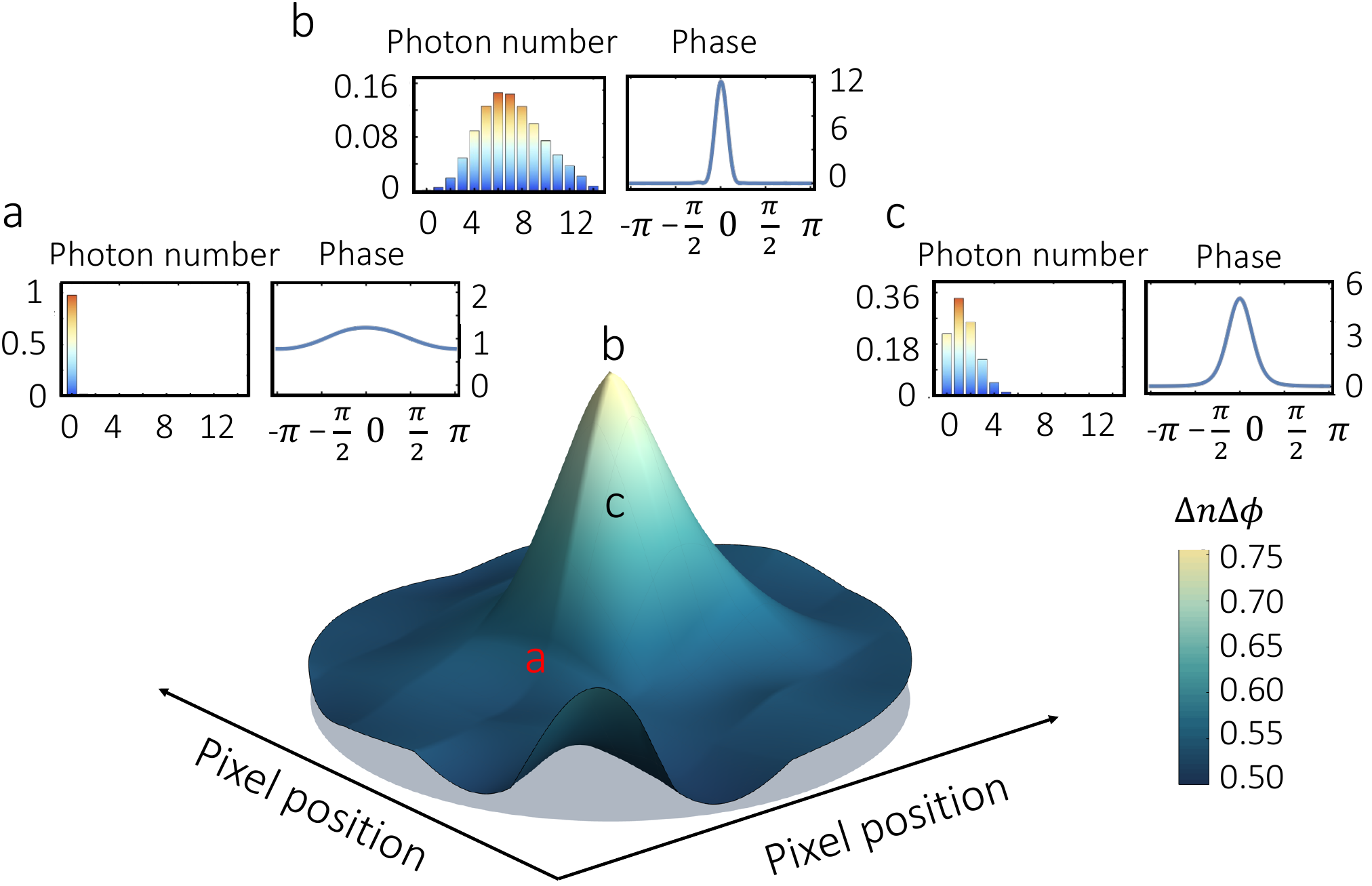}
    \caption{\textbf{Mapping the number--phase uncertainty of a multiphoton field.}
Our spatially resolved quantum state tomography enables the direct experimental characterization of the number--phase uncertainty product, $\Delta n\Delta\phi$, across a coherent multiphoton field. The measured uncertainty exhibits a pronounced spatial dependence, approaching the lower bound in regions of low photon number and increasing toward the center of the field. The photon-number and phase distributions at representative positions \textbf{a--c} reveal the complementary fluctuations underlying this behavior. In \textbf{a}, the low photon-number region is dominated by the vacuum contribution and accompanied by a broad phase distribution. Toward the center of the field, as illustrated in \textbf{b} and \textbf{c}, the photon-number distribution broadens as higher photon-number states become populated, while the complementary phase distribution becomes increasingly localized around the coherent-state phase. These measurements reveal the reciprocal evolution of photon-number and phase fluctuations across a spatially varying multiphoton field, providing a direct experimental realization of the fundamental number--phase uncertainty relation.}
    \label{fig:4}
\end{figure}

Having experimentally established the Fourier duality between photon number and phase, we finally investigate the uncertainty relation governing these complementary quantities \cite{PeggJMO1990}. Using the local quantum states reconstructed through our tomography, we evaluate the number--phase uncertainty relation bounded by Eq.~(\ref{Eq2}) at each transverse position of the multiphoton field. The calculation of this product is discussed in section 2 of the SI. Figure~\ref{fig:4} shows the resulting spatial distribution of the uncertainty product $\Delta n\Delta\phi$. Remarkably, the uncertainty exhibits a pronounced spatial dependence, approaching its lower bound in the low-photon-number regions of the field and increasing toward the beam center. This behavior directly reflects the state-dependent character of Eq.~(\ref{Eq2}), for which the uncertainty is governed by the local photon-number fluctuations and the phase information encoded in the coherences of the reconstructed quantum state. The insets in Figure~\ref{fig:4}\textbf{a--c} reveal the physical origin of this behavior through the photon-number and phase distributions at representative positions. In the low-intensity region, the field is dominated by the vacuum contribution and exhibits a broad phase distribution. As the mean photon number increases toward the center of the field, progressively higher photon-number states contribute to the multiphoton system, broadening the photon-number distribution while the complementary phase distribution becomes increasingly localized around the coherent-state phase. These measurements directly expose the reciprocal fluctuations of particle number and quantum phase across the optical field. To our knowledge, they constitute the first experimental realization of the number--phase uncertainty relation for multiphoton quantum phase states, providing experimental access to a fundamental relation that has remained inaccessible despite decades of theoretical investigation of quantum phase \cite{CarruthersRMP1968, SpechtNP2009, ChrapkiewiczNatPhoton2016, PeggJMO1990, SusskindPRF1964, LynchPR1995, NohPRL1991}.

We have established an experimental framework for measuring the quantum phase of multiparticle systems, providing access to a physical quantity that has remained elusive throughout the history of quantum mechanics \cite{CarruthersRMP1968, SpechtNP2009, ChrapkiewiczNatPhoton2016, PeggJMO1990, SusskindPRF1964, LynchPR1995, NohPRL1991}. By combining homodyne interferometry with photon-number-resolving detection \cite{SmitheyPRL1993, VogelPRA1989, YouAPR2020}, we reconstructed complementary photon-number and phase representations of multiphoton fields containing up to fifteen particles and directly characterized their number--phase uncertainty relation. Beyond measuring quantum phase, our framework enables the projection of multiphoton fields onto states with well-defined photon number or quantum phase, extending the powerful concept of photon-number post-selection to its complementary phase basis. We further demonstrated the spatially resolved Fourier duality connecting these representations, revealing how multiphoton coherences give rise to phase states and providing experimental access to mutually unbiased photon-number and phase bases \cite{AdamsonPRL2010, WoottersAnnPhys1989, Wiesniak2011, PeggJMO1990}. These capabilities establish quantum phase as an experimentally accessible resource for exploring multiparticle coherence, complementarity, and uncertainty. More broadly, phase-selective measurements offer a new degree of freedom for uncovering multiparticle phenomena that remain hidden in the photon-number representation and for manipulating coherent quantum systems. Our results thus transform the number--phase problem from a longstanding conceptual question into an experimentally accessible framework, opening new directions for fundamental quantum physics and multiparticle quantum information science \cite{ BhusalNPJ2022, MostafaviAPR2025, HongAdvSci2026, ZhangAOP2024, AspuruGuzikNatPhys2012, WalmsleyOpticaQ2023, AmicoRMP2022}.

\section{Methods}

\renewcommand{\theequation}{M\arabic{equation}}
\setcounter{equation}{0}

\noindent\textbf{Photon-counting homodyne quantum state tomography.}
We reconstruct the quantum state of the optical field using photon-counting homodyne tomography, in which photon-number-resolving detection provides the statistical frequencies of photon-counting events for different configurations of the local oscillator. These measurements are used to reconstruct the density matrix of the field through a maximum-likelihood estimation procedure. Because the optical field occupies an infinite-dimensional Hilbert space, complete reconstruction would formally require an infinite number of measurements. Experimentally, we restrict the reconstruction to a finite-dimensional Fock basis, choosing a sufficiently large photon-number cutoff to capture the relevant populations and coherences while minimizing truncation effects.

Rather than measuring a phase operator directly, we reconstruct the density matrix and subsequently evaluate the expectation values of the relevant phase observables. This approach provides access to complementary photon-number and phase representations through the same experimentally reconstructed quantum state. The mathematical construction of the phase observables and the corresponding number--phase uncertainty relation are detailed in Section 2 of the Supplementary Information.

\medskip
\noindent\textbf{Maximum-likelihood density-matrix reconstruction.}
The measured photon-counting probabilities are related to the density matrix $\hat{\rho}$ through the Born rule,
\begin{equation}
p_N = \operatorname{Tr}\left(
\hat{\Gamma}_N \hat{\rho}
\right),
\end{equation}
where $\hat{\Gamma}_N$ denotes the POVM element associated with the corresponding photon-counting outcome. Direct inversion of the measured probabilities is generally unsuitable in the presence of experimental noise, as it can produce density matrices that violate Hermiticity, positivity, or unit trace. We therefore employ the iterative $R\rho R$ maximum-likelihood algorithm to obtain a physical density matrix consistent with the measured photon-counting statistics. The reconstruction is performed through successive matrix operations, with the detailed numerical procedure described in Section 1 of the Supplementary Information.

\medskip
\noindent\textbf{Experimental coherence and detection model.}
Imperfect interference between the signal and local-oscillator fields can reduce the experimentally accessible coherence. Such effects arise from finite coherence length, spatial-mode mismatch, imperfect alignment, and other experimental imperfections. We incorporate these contributions into the POVM model by representing the optical field as a coherent component accompanied by an incoherent mode,
\begin{equation}
\hat{a}_{\mathrm{coh}}
\rightarrow
V\hat{a}_{\mathrm{coh}}
+
\sqrt{1-V^2}\hat{a}_{\mathrm{incoh}},
\end{equation}
where $V$ describes the effective coherence of the interference process. We determine this parameter from the ratio
\begin{equation}
V =
\frac{g_{\mathrm{measured}}^{(1)}}
{g_{\mathrm{theory}}^{(1)}},
\end{equation}
with $g_{\mathrm{measured}}^{(1)}$ denoting the experimentally determined first-order coherence and $g_{\mathrm{theory}}^{(1)}$ its expected value in the absence of these imperfections. The parameter can be calibrated using a coherent optical field. Including this contribution in the POVM model allows the reconstruction to account for reduced experimental coherence. As $V$ decreases, however, increasingly high signal-to-noise ratios are required for reliable reconstruction, and the reconstruction becomes uninformative in the limit of vanishing interference coherence.

\medskip
\noindent\textbf{Spatially resolved reconstruction of multiphoton states.}
We apply the tomographic procedure to optical fields with different photon-number statistics, including partially thermal fields used to investigate the complementary number and phase representations. For the spatially resolved measurements, we reconstruct the local density matrix of a coherent optical field at different transverse positions. The reconstructed density matrices provide the photon-number populations and off-diagonal coherences required to evaluate the corresponding local photon-number and phase subsystems.

The spatial variation of the detected photon number produces position-dependent signal-to-noise ratios. Regions near the center of the optical field generally provide stronger signals and more reliable density-matrix reconstructions, whereas the reduced photon-counting statistics toward the beam edges increase sensitivity to experimental noise. We characterize reconstruction quality through the fidelity with the corresponding coherent state, distinguishing this statistical measure from the physical number--phase uncertainty product evaluated in Fig.~4. Because the phase of the coherent optical field may vary across transverse positions, the local reference phase is determined from the argument of the reconstructed expectation value of the exponential phase operator, $\arg\langle\hat{E}\rangle$.

\bibliography{main}


\onecolumngrid
\clearpage

\nolinenumbers

\setcounter{section}{0}
\setcounter{subsection}{0}
\setcounter{equation}{0}
\setcounter{figure}{0}
\setcounter{table}{0}

\renewcommand{\theequation}{S\arabic{equation}}
\renewcommand{\thefigure}{S\arabic{figure}}
\renewcommand{\thetable}{S\arabic{table}}

\def\theHsection{SI.\arabic{section}}
\def\theHsubsection{SI.\arabic{section}.\arabic{subsection}}
\def\theHequation{SI.\arabic{equation}}
\def\theHfigure{SI.\arabic{figure}}
\def\theHtable{SI.\arabic{table}}


\begin{center}
{\Large\bfseries Supplementary Information:
Measuring the Quantum Phase of Multiphoton Systems\par}

\vspace{1em}

Ian Baum, Riley B. Dawkins,
Jannatul Ferdous, and Omar S. Maga\~na-Loaiza

\vspace{0.5em}

Quantum Photonics Laboratory,
Department of Physics \& Astronomy,
Louisiana State University,
Baton Rouge, LA 70803, USA
\end{center}

\vspace{1em}

This document provides supplementary information for “Measuring the Quantum Phase of Multiphoton Systems.”We provide here the derivations for all formulas used in the main paper, as well as some further discussion of their interesting features. We also provide details of our experiment.

\section{Section 1: Phase Measurements in Quantum Mechanics}

The state, $\hat\rho '$, as seen by a photon number resolving detector is sampled over multiple subsystems such as horizontal polarizaiton, vertical polarization, an incoherent mode in each polarization, and a low mean, independent Poissonian dark count subsystem. Since the photons are observed simultaneously in the detection window, different combinations of events may allow the same measured number of photons. This can be written in the Born rule as    
\begin{equation}
    p(N) = \sum_{i,j,k,l,d} p(i,j,k,l,d|i+j+k+l+d=N)
        = \sum_{i+j+k+l+d=N} Tr\{ \bigotimes_{r=1}^{5}|n_r\rangle \langle n_r| \hat\rho'\}
\end{equation}
The density matrix $\hat\rho '$ is a unitary transformation $\hat{U}$ from its initial state $\hat\rho$. The probability becomes
\begin{equation}
     Tr\{ (\sum_{i+j+k+l+d=N}\bigotimes_{r=1}^{5}|n_r\rangle \langle n_r| )\hat\rho'\} = Tr\{ \hat\Pi_N \hat\rho' \} = Tr\{ \hat\Pi_N \hat U\hat\rho \hat U^\dagger \},
\end{equation}
where the Fock projection $\hat\Pi_N$ onto these modes can be written as an integral
\begin{equation}
    \hat\Pi_N = \frac{1}{2\pi}\int_0^{2\pi}d\theta e^{i\theta(\hat{N}-N)},
\end{equation}
and $\hat{N}$ is the sum over the number operators in each detected subsystem including the dark count subsystem $\hat d$:
\begin{equation}
    \hat{N} = \vec{a}'^\dagger \vec{a}' + \hat d^\dagger \hat d.
\end{equation}
We use the cyclic properties of the trace to observe the change on our projection instead of the state.
\begin{equation}
p(N) = Tr\{ \hat U^\dagger \hat\Pi_N \hat U \hat\rho \}
\end{equation}
The unitary transformation $\hat U$ can be viewed as an operation on the annihilation operators via a unitary $\mathcal{U}$.
\begin{equation}
\begin{aligned}
    \begin{pmatrix}
    \vec{a}' \\
    \vec{e}'
    \end{pmatrix}
    &=\mathcal{U}    \begin{pmatrix}
    \vec{a} \\
    \vec{e}
    \end{pmatrix},
    &\mathcal{U}=
    \begin{pmatrix}
    J_{eff} & L \\
    K & H
    \end{pmatrix}    
\end{aligned}
\end{equation}
The detector does not see $\vec{e}'$ so we are only interested in $\hat d$ and
\begin{equation}
\label{a after unitary}
    \vec{a}' = J_{eff}\vec{a}+L\vec{e}
\end{equation}
The matrix $\mathcal{U}$ must be unitary so we assert the necessary conditions on L, K, and H to make it unitary. In general, $J_{eff}^\dagger J_{eff} \leq I$ so making $\mathcal{U}$ unitary is always possible. $J_{eff}$ is a the effective 4x4 Jones matrix on our system acting on the four modes: horizontal, vertical, incoherent horizontal, and incoherent vertical. Coherence loss will be modeled as a population of the incoherent modes also seen by the detector. It is equivalent to a population transfer from one or both the coherent modes to the incoherent modes, but here, we choose to have a transfer from the vertical mode to the incoherent vertical mode. That is, $a_V\rightarrow Va_V+\sqrt{1+V^2}a_{incoh}$. The effective Jones matrix is then related to the Jones matrix of the setup J as
\begin{equation}
\begin{aligned}
    &J_{eff}=
    \begin{pmatrix}
    JV_1 & 0 \\
    JV_2 & 0
    \end{pmatrix},&
\end{aligned}
\end{equation}
where $V_1 = diag(1,V)$, $V_2 = diag(0,\sqrt{1-V^2})$, and V is the amplitude fraction that moves to the incoherent mode $V = g^{(1)}_{measured}/g^{(1)}_{theory}$. To proceed, we normal order, 
\begin{equation}
    \hat U^\dagger \hat\Pi_n \hat U = \frac{1}{2\pi}\int_0^{2\pi}d\theta e^{-i\theta N} e^{i\theta(\vec{a'}^\dagger \vec{a'}+d^\dagger d)} = :\frac{1}{2\pi}\int_0^{2\pi}d\theta e^{-i\theta N}exp({(e^{i\theta}-1)(\vec{a'}^\dagger \vec{a'}+d^\dagger d)}) :.
\end{equation}
From expression \ref{a after unitary},
\begin{equation}
\label{n after unitary}
    \vec{a'}^\dagger \vec{a'} = (\vec{a}^\dagger J_{eff}^\dagger J_{eff} \vec{a}+\vec{a}^\dagger J_{eff}^\dagger L \vec{e}+\vec{e}^\dagger L^\dagger J_{eff} \vec{a}+\vec{e}^\dagger L^\dagger L \vec{e}).
\end{equation}
In all cases, terms containing $\vec{e}$ or $\vec{e}^\dagger$ go to zero since $\hat{e}_i\hat\rho=0 $ and $ \hat\rho \hat{e}_i^\dagger =0 $. We define the upper left corner of $J_{eff}^\dagger J_{eff}$ as M. The rest of the matrix elements are zero so this simplifies to a single subsystem. $M=V_1^\dagger J^\dagger J V_1 + V_2^\dagger J^\dagger J V_2$.
\begin{equation}
    \vec{a}^\dagger J_{eff}^\dagger J_{eff}\vec{a}=M_{HH}\hat{a}_H^\dagger\hat{a}_H + M_{HV}\hat{a}_H^\dagger\hat{a}_V+ M_{VH}\hat{a}_V^\dagger\hat{a}_H +M_{VV}\hat{a}_V^\dagger\hat{a}_V
\end{equation}
The LO is a vertically polarized coherent state $|\beta\rangle \langle \beta |$ so $\hat{a}_V\hat\rho = \beta\hat\rho$. The initial state is an arbitrary state in the horizontal polarization and vacuum in the m other subsystems. We split the horizontal portion as
\begin{equation}
    \hat\rho = \hat\rho_{sig} \otimes |\beta\rangle\langle \beta|\otimes |0\rangle\langle 0|^{\otimes m}=(I\otimes |\beta\rangle\langle \beta| \otimes|0\rangle\langle 0|^{\otimes m})(\hat\rho_{sig}\otimes I^{\otimes m+1}).
\end{equation}
Inserting this into our probability, we have
\begin{equation}
p(N) = Tr\{ \hat U^\dagger \hat\Pi_n \hat U (I\otimes |\beta\rangle\langle \beta| \otimes |0\rangle\langle 0|^{\otimes m})(\hat\rho_{sig}\otimes I^{\otimes m+1})\},
\end{equation}
and using the identity
\begin{equation}
    Tr\{R\otimes I^{\otimes n} \cdot A\}   =   Tr_1\{R\cdot Tr_{2,3,...n}\{A\}\},
\end{equation}
we immediately see that the action of the experiment can be represented as POVMs
\begin{equation}
    p(N)=Tr\{ \hat\Gamma_N \cdot\hat\rho \}.
\end{equation}
The POVM elements $\hat\Gamma_N$ are
\begin{equation}
    \hat\Gamma_N= Tr_{2,3...}\{\hat U^\dagger \hat\Pi_n \hat U (I\otimes |\beta\rangle\langle \beta| \otimes |0\rangle\langle 0|^{\otimes m})\}.
\end{equation}
We've replaced all annihilation operators with their eigenvalues since our operator is normal ordered. All subsystems but one will trace out to unity since the states are normalized. The dark count subsystem is treated as a coherent state since coherent states have Poissonian distributions. Because of this, the $d^\dagger d$ term can be replaced with the dark count mean $\bar{n}_d$. With only one untraced subsystem left, we drop the annihilation operator subscripts for brevity.
\begin{equation}
    \hat\Gamma_N = :\frac{1}{2\pi}\int_0^{2\pi}d\theta e^{-i\theta N}exp({(e^{i\theta}-1)(M_{HH}\hat{a}^\dagger\hat{a} + M_{HV}\hat{a}^\dagger\beta+ M_{VH}\beta^\dagger\hat{a} +M_{VV}|\beta|^2}+\bar{n}_d) :
\end{equation}
From here, we revert the normal ordering by using the identity
\begin{equation}
    :exp(A\hat a^{\dagger}\hat a + B\hat a^\dagger + C\hat a+D):=exp(D)exp(B\hat a^\dagger)exp(ln(1+A)\hat a^\dagger \hat a)exp(C\hat a).
\end{equation}
Since coherent states form an over-complete basis, this identity is readily proven by showing for any coherent states $\beta$ and $\alpha$, $\langle \beta|LHS|\alpha\rangle=\langle \beta|RHS|\alpha\rangle=exp(A\beta^*\alpha+B\beta^*+C\alpha+D)$. 
We get the POVM elements,
\begin{equation}
    \hat\Gamma_N=\frac{1}{2\pi}\int_0^{2\pi}d\theta e^{-i\theta N} exp(D(\theta))exp(B(\theta)a^\dagger)exp(ln(1+A(\theta))a^\dagger a)exp(C(\theta)a),
\end{equation}
where $A(\theta) = (e^{i\theta}-1)M_{HH}$, $B(\theta) = (e^{i\theta}-1)\beta M_{HV}$, $C(\theta) = (e^{i\theta}-1)\beta^* M_{VH}$, and $D(\theta) = (e^{i\theta}-1)(|\beta|^2 M_{VV}+\bar{n}_d)$. As required for POVMs, $\sum \hat\Gamma_N^\dagger \hat\Gamma_N=I$. We can't compute matrices of infinite size so the state, as well as our POVM elements, will be truncated to a dimensionality of K. For a truncated matrix, our POVM elements can be written as a sum. The chosen dimensionality is greater than the highest nonzero probability in the measured histograms. The sum runs up to $K_\theta-1$ which can be set arbitrarily high but for measured data, setting $K=K_\theta>N_{max}$ is sufficient to machine precision. We have,
\begin{equation}
    \hat\Gamma_N=\frac{1}{K_\theta}\sum_{k=0}^{K_\theta-1}e^{-i\theta_k N} exp(D(\theta_k))exp(B(\theta_k)a^\dagger)exp(ln(1+A(\theta_k))a^\dagger a)exp(C(\theta_k)a)
\end{equation}
where $\theta_k=\frac{2\pi k}{K_\theta}$. At this point, the matrix is readily calculable. This is then the expression used in the iterative maximum likelihood algorithm called the iterative-$R\rho R$ algorithm which is an iterative solution to equation 1 in the main text that is been proven to always move into the direction of higher likelihood.

The initial guess is arbitrary so we start from the maximally mixed state, $\hat\rho_0=I/K$, the density matrix at iteration $k+1$ is given by 

\begin{equation}
\hat\rho_{k+1} =\frac{(I+\epsilon \hat R_k)\hat\rho_k(I+\epsilon \hat R_k)}{\operatorname{Tr}\{(I+\epsilon \hat R_k)\hat\rho_k(I+ \epsilon \hat R_k)\}},
\end{equation}             
where $\hat R_k$ is an operator defined from the POVM elements and measurement frequencies from all settings

\begin{equation}
\hat R_k=\sum_{\vec{\theta},n}\frac{N_n(\vec{\theta})}{p_n^{(k)}(\vec{\theta})}\hat \Gamma_n(\vec{\theta}),
\qquad
p_n^{(k)}(\vec{\theta})
=
\operatorname{Tr}
[\hat\rho_k \hat\Gamma_n(\vec{\theta})].
\end{equation}
The dilution parameter $\epsilon$ prevents cycling between successive iterations. The optimal value for $\epsilon$ in the k-th iteration can be computationally expensive but having it monotonically decrease slow enough is sufficient to machine precision.

\section{Section 2: The Phase Operator}

The first question we must ask ourselves when discussing the `phase operator' is what we want to measure. One option is the geometric phase at a given point in space, but measuring geometric phase only really makes sense when dealing with a single-mode excitation of the electromagnetic field, because it is a classical property of light and not a quantum one. What we are interested in is a quantum phase, something related to the statistics of the light-source and not just its excitation mode. Unfortunately, quantum mechanics as we know it is only able to represent observables with `topologically-flat' data distributions, and phase is a circular quantity (i.e., $2\pi$ radians is the same as $0$ radians). As such, a good phase operator should \textit{correspond} to measurements of phase, but its observable data should be flat. The most promising option for this would be an operator whose eigenvalues are given by all possible $e^{i\phi}$ terms, since measurements of this operator would directly correspond to each possible phase $\phi$. In the single-mode case, it turns out that such an operator is given by
\begin{equation}
    \hat{E} = \sum_{n=0}^\infty \ket{n}\bra{n+1},
\end{equation}
with corresponding eigenvectors
\begin{equation}
    \ket{\phi} = \sum_{n=0}^\infty e^{in\phi}\ket{n}.
\end{equation}
The eigenvalue-eigenvector relation for this operator and these states is $\hat{E}\ket{\phi} = e^{i\phi}\ket{\phi}$, and this is what we were looking for. As such, we will refer to $\hat{E},\hat{E}^\dagger,$ and certain operators built from them as `phase operators' throughout this manuscript, and we will also refer to the eigenvectors $\ket{\phi}$ as `phase-vectors.' One might then conclude that all of the phase-information for a given quantum state is contained in the coefficients of its $\ket{n}\bra{n+1}$-terms, but this is only true for the first moment of our phase operator $\hat{E}$. Specifically, we have that
\begin{equation}
    \hat{E}^m = \sum_{n=0}^\infty \ket{n}\bra{n+m}.
\end{equation}
Furthermore, the information corresponding to the expectation-value of $\hat{E}^{\dagger m}$ will be contained in the coefficients of a density matrix's $\ket{n+m}\bra{n}$-terms. So, while the diagonal elements of a given density matrix in the Fock basis are irrelevant to measuring the state's phase information, it is also the case that all off-diagonal elements are relevant, and not just a select-few. This is in contrast to moments of the number operator, which depend wholly on the \textit{diagonal} elements of a density matrix's Fock-representation. In this sense, number and phase measurements probe two distinct parts of a given quantum state. Now, we can also write Fock states in terms of phase eigenvectors via the relation
\begin{equation}
    \ket{n} = \frac{1}{2\pi}\int d\phi e^{-in\phi}\ket{\phi},
\end{equation}
where here it is understood that any integral over $\phi$ is taken over some arbitrary interval of length $2\pi$. To prove this, we show that
\begin{equation}
    \begin{aligned}
        \bra{m}\left(\frac{1}{2\pi}\int d\phi e^{-in\phi}\ket{\phi}\right) &= \frac{1}{2\pi}\int d\phi e^{-in\phi}\braket{m|\phi}\\
        &= \frac{1}{2\pi}\int d\phi e^{-in\phi}e^{im\phi}\\
        &= \frac{1}{2\pi}(2\pi \delta_{nm})\\
        &= \delta_{nm}.
    \end{aligned}
\end{equation}
This tells us that the Fock-basis representation of the integral over $\ket{\phi}$ vectors is given by $\ket{n}$, as desired. From this, we see that Fock-vectors and phase-vectors are two conjugate bases of the photonic Hilbert-space, where the phase-vector basis is overcomplete as opposed to being orthonormal. As such, transforming between these bases requires an integral in one direction, and a sum in the other. Because phase vectors and Fock vectors form conjugate bases under the above discrete-continuous Fourier transformation relation, we should expect to see an interesting Heisenberg uncertainty principle between measurements of photon-number and measurements of phase. However, $\hat{E}$ is notably not Hermitian. To fix this issue, we introduce the `trigonometric' phase operators
\begin{equation}
    \begin{aligned}
        \hat{C} &= \frac{1}{2}(\hat{E} + \hat{E}^\dagger),\\
        \hat{S} &= \frac{1}{2i}(\hat{E} - \hat{E}^\dagger).
    \end{aligned}
\end{equation}
The operator $\hat{C}$ corresponds to the phase's `cosine,' and the operator $\hat{S}$ corresponds to its `sine.' By measuring both of these operators separately, we can obtain the desired expectation value $\braket{\hat{E}} = \braket{\hat{C}} + i \braket{\hat{S}}$. The important commutation relations between these phase-operators are given by
\begin{equation}
    \begin{aligned}
        [\hat{E},\hat{E}^\dagger] &= \sum_{n=0}^\infty \ket{n}\bra{n} - \sum_{n=1}^\infty \ket{n}\bra{n} = \ket{0}\bra{0},\\
        [\hat{C},\hat{S}] &= \frac{1}{4i}(-[\hat{E},\hat{E}^\dagger] + [\hat{E}^\dagger,\hat{E}]) = \frac{i}{2}\ket{0}\bra{0}.
    \end{aligned}
\end{equation}
Phase operators also obey interesting commutation relations with the number operator, $\hat{n} = \sum_{n=0}^\infty n \ket{n}\bra{n}$. In particular,
\begin{equation}
    \begin{aligned}
        [\hat{n},\hat{E}] &= \sum_{n=0}^\infty n\ket{n}\bra{n+1} - \sum_{n=0}^\infty (n+1)\ket{n}\bra{n+1} = -\sum_{n=0}^\infty \ket{n}\bra{n+1} = -\hat{E},\\
        [\hat{n},\hat{E}^\dagger] &= \sum_{n=0}^\infty (n+1)\ket{n+1}\bra{n} - \sum_{n=0}^\infty n\ket{n+1}\bra{n} = \sum_{n=0}^\infty \ket{n+1}\bra{n} =\hat{E}^\dagger.
    \end{aligned}
\end{equation}
Furthermore, these commutation relations can be implemented to derive
\begin{equation}
    \begin{aligned}
        [\hat{n},\hat{C}] &= \frac{1}{2}(-\hat{E}+\hat{E}^\dagger) = -i\frac{1}{2i}(\hat{E}-\hat{E}^\dagger) = -i\hat{S},\\
        [\hat{n},\hat{S}] &= \frac{1}{2i}(-\hat{E}-\hat{E}^\dagger) = i\frac{1}{2}(\hat{E}+\hat{E}^\dagger) = i\hat{C}.
    \end{aligned}
\end{equation}
With these, we are able to obtain the following Heisenberg uncertainty principals:
\begin{equation}
    \begin{aligned}
        \Delta n \Delta C &\geq \frac{1}{2}|\braket{[\hat{n},\hat{C}]}| = \frac{1}{2}|\braket{\hat{S}}|,\\
    \Delta n \Delta S &\geq \frac{1}{2}|\braket{[\hat{n},\hat{S}]}| = \frac{1}{2}|\braket{\hat{C}}|.
    \end{aligned}
\end{equation}
The downside to these equations is that the right-hand-side depends on the state, so they are not as simple as, say, the principle relating position and momentum observables ($\Delta x \Delta p \geq 1/2$). If one wants to compute the phase itself, $\phi$, then they would calculate $\phi = \tan^{-1}(\braket{\hat{S}}/\braket{\hat{C}})$. In other words, the phase is not a directly-observable quantity, and so things like the uncertainty $\Delta\phi$ must be inferred from measurements of $\hat{C}$ and $\hat{S}$ through error-propagation. Indeed, even if we had a direct operator for $\phi$, we would not know how to describe uncertainty in its measurement due to the circular quality of phase data. Applying standard error-propagation, we can compute the formula for phase-uncertainty via
\begin{equation}
    \begin{aligned}
        \Delta\phi &= \Delta\left(\tan^{-1}(S/C)\right)\\
        &= \sqrt{\left(\frac{\partial(\tan^{-1}(S/C))}{\partial S}\Delta S\right)^2 + \left(\frac{\partial(\tan^{-1}(S/C))}{\partial C}\Delta C\right)^2 + 2\frac{\partial(\tan^{-1}(S/C))}{\partial S}\frac{\partial(\tan^{-1}(S/C))}{\partial C}\text{Cov}(S,C)}\\
        &= \sqrt{\left(\frac{C}{C^2+S^2}\Delta S\right)^2 + \left(-\frac{S}{C^2+S^2}\Delta C\right)^2 - 2\frac{S C}{(S^2 + C^2)^2}\text{Cov}(S,C)}\\
        &= \frac{\sqrt{S^2 (\Delta C)^2 + C^2 (\Delta S)^2 - 2 S C \text{Cov}(S,C)}}{S^2 + C^2}.
    \end{aligned}
\end{equation}
Classically, the covariance term $\text{Cov}(S,C)$ is given by $\braket{SC} - \braket{S}\braket{C}$, but in terms of quantum-mechanical measurements this term would be given by $\text{Cov}(S,C) = \braket{\hat{S}\hat{C}+\hat{C}\hat{S}}/2 - \braket{\hat{S}}\braket{\hat{C}}$. As such, we wind up with the photon-number-phase uncertainty relation
\begin{equation}
\begin{aligned}
    \Delta n \Delta \phi &= \Delta n \frac{\sqrt{\braket{\hat{S}}^2 (\Delta C)^2 + \braket{\hat{C}}^2 (\Delta S)^2- 2\braket{\hat{C}}\braket{\hat{S}}\text{Cov}(S,C)}}{\braket{\hat{S}}^2 + \braket{\hat{C}}^2}\\
    &= \frac{\sqrt{\braket{\hat{S}}^2 (\Delta n\Delta C)^2 + \braket{\hat{C}}^2 (\Delta n\Delta S)^2- 2(\Delta n)^2\braket{\hat{C}}\braket{\hat{S}}\text{Cov}(S,C)}}{\braket{\hat{S}}^2 + \braket{\hat{C}}^2}
    \\&\geq \frac{\sqrt{\frac{1}{4}\braket{\hat{S}}^4 + \frac{1}{4}\braket{\hat{C}}^4 - 2 (\Delta n)^2\braket{\hat{C}}\braket{\hat{S}}\text{Cov}(S,C)}}{\braket{\hat{C}}^2+\braket{\hat{S}}^2}.
\end{aligned}
\end{equation}
The behavior of this uncertainty-principle is strongly dependent on the quantity $\mathcal{P}(\hat{\rho}) = \braket{\hat{C}}\braket{\hat{S}}\text{Cov}(S,C)$, which can range from $-\Delta C \Delta S|\braket{\hat{C}}\braket{\hat{S}}|$ to $\Delta C \Delta S|\braket{\hat{C}}\braket{\hat{S}}|$ because of the Cauchy-Schwarz inequality. When $\mathcal{P}(\hat{\rho}) = -\Delta C \Delta S|\braket{\hat{C}}\braket{\hat{S}}|$, the right-hand-side simplifies to $1/2$, which is nice because it calls back to the position-momentum uncertainty principle (though this is notably unitless). However, this is an upper-bound on the uncertainty principle's lower-bound, and notably there is the possibility of an uncertainty principle $\Delta n \Delta\phi \geq 0$, suggesting that some states will have no minimum bound on the right-hand-side of the uncertainty principle, under the right conditions. As noted above, these measurements of phase are dependent solely on the off-diagonal elements of the density matrix. As such, by measuring all the components of a given density matrix through state tomography, we can decompose a state not only into its photon-number subsystems, but also into its phase-subsystems, the distributions which determine the statistical phase of the state. In the next section, we will discuss this process in detail.

We have introduced the phase operators $\hat{E},\hat{C},$ and $\hat{S}$. Furthermore, we represented them in the Fock (or photon-number) basis. In this basis, it was apparent that the information about a state's phase was contained entirely in its off-diagonal components. Conversely, information about a state's photon-number distribution is contained only in the diagonal components. In photonics, one typically defines multiphoton subsystems via the probability of observing a particular number of photons (i.e., a measurement of a particular multiphoton wavepacket). Conversely, one might then aim to define a phase subsystem as the probability of observing a particular phase values (which we could then also refer to as a phase-wavepacket). Indeed, extracting these subsystems would potentially allow for a more intuitive characterization of phase than that which is obtained through macroscopic measurements of the sine and cosine operators, and as such it is one of the fundamental goals of this manuscript. In this section, we will discuss the theory behind their extraction. However, it is unclear in the Fock-basis how the off-diagonal components can be used to determine a particular probability for observing a particular phase value. It will therefore be useful to work in a conjugate basis where such a structure becomes apparent, and it will turn out that this basis is the overcomplete phase basis spanned by the $\ket{\phi}$ operators. To show this, we first write the identity operator in this basis as
\begin{equation}
    \begin{aligned}
        \hat{\mathbb{I}} &= \sum_{n=0}^\infty\ket{n}\bra{n}\\
        &= \sum_{n=0}^\infty\frac{1}{2\pi}\int d\phi e^{in\phi}\ket{n}\bra{\phi}\\
        &= \frac{1}{2\pi}\int d\phi \ket{\phi}\bra{\phi}.
    \end{aligned}
\end{equation}
Then, using the eigenvalue relation $\hat{E}\ket{\phi} = e^{i\phi}\ket{\phi}$, we can quickly show that
\begin{equation}
    \begin{aligned}
        \hat{E} &= \frac{1}{2\pi}\int d\phi e^{i\phi}\ket{\phi}\bra{\phi},\\
        \hat{E}^\dagger &= \frac{1}{2\pi}\int d\phi e^{-i\phi}\ket{\phi}\bra{\phi},\\
        \hat{C} &= \frac{1}{2\pi}\int d\phi \cos(\phi)\ket{\phi}\bra{\phi},\\
        \hat{S} &= \frac{1}{2\pi}\int d\phi \sin(\phi)\ket{\phi}\bra{\phi}.
    \end{aligned}
\end{equation}
This suggests that these operators represent the expectation-value of their classical counterparts over phase vectors. To make this notion more concrete, let's suppose that we have a density matrix, written in the Fock basis, given by
\begin{equation}
    \hat{\rho} = \sum_{n,m=0}^\infty \rho_{nm}\ket{n}\bra{m}.
\end{equation}
This density matrix will satisfy the condition $\sum_{n=0}^\infty \rho_{nn} = 1$. Furthermore, the diagonal elements must all be positive and we must have that $\rho_{nm} = \rho_{mn}^*$. Lastly, $\hat{\rho}$ is semi-positive definite, meaning that $\braket{\psi|\hat{\rho}|\psi}\geq 0$ for all vectors $\ket{\psi}$. We can write this operator in the phase-vector basis as
\begin{equation}
    \begin{aligned}
        \hat{\rho} &= \sum_{n,m=0}^\infty \rho_{nm}\ket{n}\bra{m}\\
        &= \sum_{n,m=0}^\infty \rho_{nm}\frac{1}{4\pi^2}\int d\phi d\phi'e^{-in\phi + im\phi'}\ket{\phi}\bra{\phi'}\\
        &= \frac{1}{2\pi}\int d\phi d\phi' \rho(\phi,\phi')\ket{\phi}\bra{\phi'},
    \end{aligned}
\end{equation}
where here we have defined
\begin{equation}
    \rho(\phi,\phi') = \frac{1}{2\pi}\sum_{n,m=0}^\infty \rho_{nm}e^{-in\phi + im\phi'}.
\end{equation}
This function notably satisfies the condition
\begin{equation}
    \int d\phi \rho(\phi,\phi) = \frac{1}{2\pi}\int d\phi \sum_{n,m=0}^\infty \rho_{nm}e^{i\phi(m-n)} = \frac{1}{2\pi}\sum_{n,m=0}^\infty \rho_{nm}(2\pi \delta_{nm}) = \sum_{n=0}^\infty \rho_{nn} = 1,
\end{equation}
as well as that
\begin{equation}
    \rho^*(\phi,\phi') = \sum_{n,m=0}^\infty \rho^*_{nm}e^{i\phi n - i\phi' m} = \sum_{n,m=0}^\infty \rho_{mn}e^{-i\phi'm + i\phi n} = \rho(\phi',\phi).
\end{equation}
However, showing that $\rho(\phi,\phi)\geq0$ for all $\phi$ is a bit more difficult. We will first show that $\text{Tr}[\hat{\rho}\hat{A}^\dagger\hat{A}]\geq0$ for all operators $\hat{A}$. To do this, we start out by noting that $\hat{A}^\dagger\hat{A}$ is Hermitian, and therefore that it has real eigenvalues. Furthermore, given a normalized eigenvector $\ket{a_i}$ with corresponding eigenvalue relation $\hat{A}^\dagger\hat{A}\ket{a_i} = a_i\ket{a_i}$, we find that
\begin{equation}
    \braket{a_i|\hat{A}^\dagger\hat{A}|a_i} = a_i = \Vert\hat{A}\ket{a_i}\Vert^2 \geq 0.
\end{equation}
So, the eigenvalues $a_i$ must all be greater-than-or-equal-to $0$. As such, we get that
\begin{equation}
    \text{Tr}[\hat{\rho}\hat{A}^\dagger\hat{A}] = \sum_i a_i \braket{a_i|\hat{\rho}|a_i} \geq 0,
\end{equation}
since we can write $\hat{A}^\dagger\hat{A} = \sum_i a_i\ket{a_i}\bra{a_i}$ and since $\hat{\rho}$ is semi-positive-definite. Finally, consider the operator
\begin{equation}
    \hat{A} = \frac{1}{\sqrt{2\pi}}\ket{0}\bra{\phi}.
\end{equation}
This operator satisfies $\hat{A}^\dagger\hat{A} = \ket{\phi}\bra{\phi}/(2\pi)$, and more importantly it yields
\begin{equation}
    \text{Tr}[\hat{\rho}\hat{A}^\dagger\hat{A}] = \frac{1}{2\pi}\braket{\phi|\hat{\rho}|\phi} = \frac{1}{2\pi}\sum_{n,m=0}^\infty \rho_{nm}e^{i\phi(m-n)} = \rho(\phi,\phi)\geq 0,
\end{equation}
proving that $\rho(\phi,\phi)$ is a probability distribution. Next, let's study how this representation of the density matrix in terms of phase vectors corresponds to measurements of phase operators. To do this, however, we must first give a precise definition of what a phase operator is. Let's first note that one can write (by inserting the identity operator)
\begin{equation}
    \hat{E}\hat{E}^\dagger = \frac{1}{2\pi}\int d\phi e^{i\phi}\ket{\phi}\bra{\phi}e^{-i\phi} = \frac{1}{2\pi}\int d\phi \ket{\phi}\bra{\phi} = \hat{\mathbb{I}}.
\end{equation}
As such, one also obtains that $\hat{E}^\dagger \hat{E} = \ket{0}\bra{0}+\hat{\mathbb{I}}$. We might be tempted to define phase operators as those which can be written as arbitrary sums of products of the $\hat{E}$ and $\hat{E}^\dagger$ operators, but technically all operators can be written in terms of $\hat{E}$ and $\hat{E}^\dagger$ operators since
\begin{equation}
    \ket{n}\bra{m} = \hat{E}^{\dagger n}\ket{0}\bra{0}\hat{E}^m = \hat{E}^{\dagger (n+1)}\hat{E}^{m+1}-\hat{E}^{\dagger n}\hat{E}^{m}.
\end{equation}
As such, the class of operators known as `phase operators' should be much more restrictive than that which can be built out of the full algebra generated by the operators $\hat{E}$ and $\hat{E}^\dagger$. To justify the definition which we will shortly give, let's first clarify what we mean when we refer to an operator as a `number operator.' Specifically, a number operator is any operator which is diagonal in terms of Fock states, meaning operators of the form
\begin{equation}
    \hat{F}_\text{N} = \sum_{n=0}^\infty f_n \ket{n}\bra{n}.
\end{equation}
This definition is based on the fact that all operators of this form can be measured by way of a photon-number-resolving detector, since this allows for the determination of the quantum state's diagonal Fock-basis elements. Motivated by this definition, we will define phase operators as those which are diagonal in terms of phase states, meaning operators of the form
\begin{equation}
    \hat{F}_{\text{P}} = \frac{1}{2\pi}\int d\phi f(\phi)\ket{\phi}\bra{\phi}.
\end{equation}
To understand the behavior of these operators, we will first derive a helpful result:
\begin{equation}
    \begin{aligned}
        \braket{\phi|\hat{\rho}|\phi'} &= \sum_{n,m=0}^\infty \rho_{nm} \braket{\phi|n}\braket{m|\phi'} = \sum_{n,m=0}^\infty \rho_{nm} e^{-in\phi + im\phi'} = 2\pi\rho(\phi,\phi').
    \end{aligned}
\end{equation}
From this, we quickly obtain that
\begin{equation}
    \text{Tr}[\hat{\rho}\hat{F}_{\text{P}}] = \frac{1}{2\pi}\int d\phi f(\phi)\braket{\phi|\hat{\rho}|\phi} = \int d\phi f(\phi)\rho(\phi,\phi).
\end{equation}
As such, we conclude that the expectation value of a phase operator depends solely on the probability distribution $\rho(\phi,\phi)$, which we could intuit as the `diagonal part' of the phase-basis representation of $\hat{\rho}$. Furthermore, this result is the stable-ground on which we can now interpret $\rho(\phi,\phi)$ as the probability of observing the phase $\phi$. Since the phase-distribution $f(\phi)$ of the operator $\hat{F}_{\text{p}}$ is only defined over a $2\pi$-interval, its Fourier-series decomposition is well-defined, and given by
\begin{equation}
    f(\theta) = \sum_{n=-\infty}^\infty c_n e^{in\phi}.
\end{equation}
Plugging this into the operator $\hat{F}_{\text{P}}$ then yields
\begin{equation}
    \hat{F}_{\text{P}} = \frac{1}{2\pi}c_0\hat{\mathbb{I}} + \frac{1}{2\pi}\sum_{n=1}^\infty \left(c_n \hat{E}^n + c_{-n}\hat{E}^{\dagger n}\right).
\end{equation}
As a result, we conclude that all phase operators are given by sums of powers of $\hat{E}$ and $\hat{E}^\dagger$. One useful analytical trick to determine if an operator is a phase operator or not is to take its phase normal-order. This process of normal ordering, using semicolons $;\cdot ;$, involves placing all $\hat{E}^\dagger$ terms on the right and all $\hat{E}$ terms on the left. For instance, $;\hat{E}^\dagger\hat{E}; = \hat{E}\hat{E}^\dagger = \hat{\mathbb{I}}$. From the above Fourier decomposition, we can see that all phase operators are naturally phase-normal-ordered. Furthermore, since $\hat{E}\hat{E}^\dagger = \hat{\mathbb{I}}$, it is also the case that any phase-normal-ordered operator is itself a phase operator. One can therefore check if a given operator is a phase operator by taking its phase-normal-order and seeing whether or not they obtain a different operator. For instance, given the Fock projector $\ket{n}\bra{m}$, one obtains
\begin{equation}
    ;\ket{n}\bra{m}; = ;\left(\hat{E}^{\dagger (n+1)}\hat{E}^{m+1}-\hat{E}^{\dagger n}\hat{E}^{m}\right); = \hat{E}^{m+1}\hat{E}^{\dagger (n+1)}-\hat{E}^{m}\hat{E}^{\dagger n} = \hat{E}^{m}\hat{E}^{\dagger n}-\hat{E}^{m}\hat{E}^{\dagger n} = 0.
\end{equation}
As a result, all Fock projectors (even those which are off-diagonal) are not phase operators. Similarly, phase projectors of the form $\ket{\phi}\bra{\phi'}$ can be written in the Fock basis as
\begin{equation}
    \ket{\phi}\bra{\phi'} = \sum_{n,m=0}^\infty e^{i\phi n-i\phi' m}\ket{n}\bra{m},
\end{equation}
which is clearly not a number operator since it is not diagonal in the Fock basis (even when $\phi\neq \phi'$, the corresponding `off-diagonal' phase-projector case). Another interesting correspondence in this formulation is between which parts of the Fock-basis density matrix correspond to phase-operator information, and which parts of the phase-basis density matrix correspond to number-operator information. Specifically, we have that
\begin{equation}
    \text{Tr}[\hat{\rho}\ket{\phi}\bra{\phi}] = \sum_{n,m=0}^\infty \rho_{nm}e^{i\phi(m-n)} = \sum_{n\neq m} \rho_{nm}e^{i\phi(m-n)} + 1.
\end{equation}
As such, the `phase information' is contained entirely within the off-diagonal elements of the Fock basis. Of course, these density matrix elements depend partially on the diagonal elements of the density matrix since $\hat{\rho}$ must be semi-positive-definite, so number and phase are not completely independent. Conversely, we also have that
\begin{equation}
    \text{Tr}[\hat{\rho}\ket{n}\bra{n}] = \frac{1}{2\pi}\int d\phi d\phi' \rho(\phi,\phi')\braket{n|\phi}\braket{\phi'|m} = \frac{1}{2\pi}\int d\phi d\phi' \rho(\phi,\phi')e^{in(\phi-\phi')}.
\end{equation}
It is less clear in this case that the right-hand-side does not `depend' on the diagonal values $\rho(\phi,\phi)$, but this becomes apparent when one evaluates the integral as a Riemann sum, where the $\phi=\phi'$ parts of the pixelation will be irrelevant to the result regardless of how high the integral's resolution is. Of course, the off-diagonal values will actually determine the diagonal values as a result of continuity, but this is similar to how the diagonals and off-diagonals are not completely independent in the Fock-basis either. So, roughly-speaking, number operator information is contained in the diagonals of the Fock basis, and phase operator information is contained separately in the diagonals of the phase basis. However, the expectation values of these operators are inherently linked by the fact that quantum states must be semi-positive definite, and this is what is ultimately responsible for the Heisenberg uncertainty principals which relate number and phase operators. Additionally, we note that by combining number and phase operators, one can fully characterize the quantum state $\hat{\rho}$. For instance, by measuring operators of the form $\ket{n}\bra{n}\hat{E}^{(m-n)} = \ket{n}\bra{m}$, for $m\geq n$, and their Hermitian conjugates, one can completely reconstruct the density matrix in the Fock basis. However, this information is only accessible when the operator algebras are combined in such a way. Now, lastly, we will define phase subsystems in analogy to photon-number subsystems. Specifically, we had previously defined $\braket{n|\hat{\rho}|n} = \rho_{nn}$ as the $n$-photon `subsystem' for a single-mode state, and so here we will subsequently define $\braket{\phi|\hat{\rho}|\phi}/(2\pi) = \rho(\phi,\phi)$ as the $\phi$-phase `subsystem' of $\hat{\rho}$. Our use of the word subsystem will become more justified when we consider phase-measurements in the infinite-mode, field-theoretic framework for the free electromagnetic field.

\section{Section 3: Multiphoton and Phase Subsystems in Quantum Field Theory}

In the standard quantization of the electromagnetic field, one finds that plane-wave photons are admitted, and that smeared photons (those which consist of a probability distribution over plane-waves) are not. This is a result of the commutation relations between photonic creation and annihilation operators, together with the fundamental postulates of quantum mechanics which require that a state's trace be equal to $1$. Unfortunately, plane-waves represent fields with a uniform intensity distribution over all of space, and such fields are clearly unphysical. Therefore, we see that the standard quantization of light admits unphysical fields, while simultaneously refusing to admit physical fields, such as Gaussian beams, which are given by a smearing over plane-waves. This conundrum is resolved by recalling that classical electromagnetism is a field-theory, and then proceeding to quantize the electromagnetic field as a field-theory, in opposition to the standard procedure of mechanical-quantization. In this paradigm, plane-wave annihilation operators are given by $\hat{a}_s(\mathbf{k})$, which corresponds to a plane-wave of polarization $s$ and wavenumber $\mathbf{k}$. These operators satisfy the commutation relation
\begin{equation}
    [\hat{a}_{s}(\mathbf{k}),\hat{a}^\dagger_{s'}(\mathbf{k}')] = (2\pi)^3 \delta_{ss'}\delta(\mathbf{k}-\mathbf{k}').
\end{equation}
It is the presence of this Dirac-delta function which makes plane-waves unphysical and smeared-photons physical in the field-theoretic approach. One can obtain a spatial annihilation operator through Fourier transformation via
\begin{equation}
    \hat{a}_s(\mathbf{r}) = \int \frac{d^3 k}{(2\pi)^{3/2}} e^{-i\mathbf{k}\cdot\mathbf{r}}\hat{a}_s(\mathbf{k}).
\end{equation}
These operators, similar to the plane-wave annihilation operators, satisfy the commutation relation $[\hat{a}_{s}(\mathbf{r}),\hat{a}^\dagger_{s'}(\mathbf{r}')] = (2\pi)^3 \delta_{ss'}\delta(\mathbf{r}-\mathbf{r}')$. Instead of representing the annihilation of a plane-wave, they represent the annihilation of a light-source which is confined to exactly a point. These operators are perhaps more intuitive than the plane-wave annihilation operators for understanding smeared photonic modes. Specifically, a smeared annihilation operator is given by
\begin{equation}
    \hat{a}_{s,f} = \int \frac{d^3r}{(2\pi)^{3/2}}f(\mathbf{r})\hat{a}_s(\mathbf{r}).
\end{equation}
The function $f(\mathbf{r})$ is normalized such that $[\hat{a}_{s,f},\hat{a}_{s,f}^\dagger] = 1$. As a result of these commutation relations, such photons represent physical modes. It is useful to note that we can always choose some orthonormal basis of functions $\phi_i(\mathbf{r})$ which satisfies
\begin{equation}
    \sum_i \phi_i^*(\mathbf{r})\phi_i(\mathbf{r}') = \delta(\mathbf{r}-\mathbf{r}').
\end{equation}
The annihilation operators $\hat{a}_{s,\phi_i}$ which correspond to these functions then serve as an `orthonormal basis' of photonic modes, in the sense that
\begin{equation}
    \begin{aligned}
        \hat{a}_s(\mathbf{r}) &= \int d^3r' \delta(\mathbf{r}-\mathbf{r}')\hat{a}_s(\mathbf{r}')\\
        &= \int d^3r' \sum_i \phi^*_i(\mathbf{r})\phi_i(\mathbf{r}')\hat{a}_s(\mathbf{r}')\\
        &= (2\pi)^{3/2}\sum_i \phi^*_i(\mathbf{r})\hat{a}_{s,\phi_i}.
    \end{aligned}
\end{equation}
As such, we can think of the quantum electromagnetic field in terms of a countably-infinite set of orthonormal photonic modes, instead of an uncountably-infinite set of plane-waves. In this paper, we are specifically interested in the case where a quantum state contains photons of only a single mode, but the multimode case is useful to understand as we will later split this single mode into a multimode representation. States composed of $(s,f)$-photons in quantum field theory can be described via the Fock-space of vectors 
\begin{equation}
    \ket{N}_{s,f} = \frac{(\hat{a}_{s,f}^{\dagger})^N}{\sqrt{N!}}\ket{0}.
\end{equation}
Here, the vacuum vector $\ket{0}$ is defined as the unique vector which is annihilated by all $\hat{a}_s(\mathbf{k})$. Since the algebra of this single-mode subsystem is identical to that of any single-mode quantum-mechanical system, all results concerning such things as coherent states and Wigner-functions will hold. Now, things become more interesting when we consider photon-number-resolving detectors. In particular, the number operator is given by
\begin{equation}
    \hat{n} = \sum_s\int \frac{d^3 r}{(2\pi)^3} \hat{a}^\dagger_s(\mathbf{r})\hat{a}_s(\mathbf{r}) = \sum_{s,i}\hat{a}^\dagger_{s,\phi_i}\hat{a}_{s,\phi_i}.
\end{equation}
This operator counts the total number of quanta present in all modes of a given state. When restricted to the $(s,f)$-mode subspace, it is identical in behavior to the number operator $\hat{n}_{s,f} = \hat{a}^\dagger_{s,f}\hat{a}_{s,f}$. Physical detectors occupy finite regions of space, and can often be represented as detecting only a single photonic mode, which we will call the detector mode $D(\mathbf{r})$. It is the integral-product $\int d^3r D(\mathbf{r})f^*(\mathbf{r})$ which yields many of the interesting, uniquely field-theoretic effects in quantum photonics. One important example of a photon detector is the point-detector, a detector of small, finite volume such that the change in all $\phi_i(\mathbf{r})$ is sufficiently small over the volume of the detector. The number operator which corresponds to this detector is given by
\begin{equation}
    \begin{aligned}
        \hat{n}_{s,V}(\mathbf{r}) =& \int_{B(\mathbf{r},V)} \frac{d^3r'}{(2\pi)^3}\hat{a}^\dagger_s(\mathbf{r}')\hat{a}_s(\mathbf{r}')\\
        =& \sum_{i,j}\left(\int_{B(\mathbf{r},V)} d^3r' \phi_i(\mathbf{r}')\phi_j^*(\mathbf{r}')\right)\hat{a}^\dagger_{s,\phi_i}\hat{a}_{s,\phi_j}\\
        \approx&V(2\pi)^3\sum_{i,j}\phi_i(\mathbf{r})\phi_j^*(\mathbf{r})\hat{a}^\dagger_{s,\phi_i}\hat{a}_{s,\phi_j}\\
        \approx& \frac{1}{V(2\pi)^3}\left(\sum_i\int_{B(\mathbf{r},V)} d^3r' (2\pi)^{3/2}\phi_i(\mathbf{r}')\hat{a}^\dagger_{s,\phi_i}\right)\\&\times\left(\sum_j\int_{B(\mathbf{r},V)} d^3r'' (2\pi)^{3/2}\phi_j^*(\mathbf{r}'')\hat{a}_{s,\phi_j}\right)\\
        =& \left(\frac{1}{\sqrt{V}}\int_{B(\mathbf{r},V)}\frac{d^3r'}{(2\pi)^{3/2}}\hat{a}^\dagger_s(\mathbf{r}')\right)\left(\frac{1}{\sqrt{V}}\int_{B(\mathbf{r},V)} \frac{d^3r''}{(2\pi)^{3/2}}\hat{a}_s(\mathbf{r}'')\right)\\
        \equiv& \hat{a}_{s,V}^\dagger(\mathbf{r})\hat{a}_{s,V}(\mathbf{r}),
    \end{aligned}
\end{equation}
where here we have defined the normalized annihilation operator
\begin{equation}
    \hat{a}_{s,V}(\mathbf{r}) = \frac{1}{\sqrt{V}}\int_{B(\mathbf{r},V)} \frac{d^3r'}{(2\pi)^{3/2}}\hat{a}_s(\mathbf{r}'),
\end{equation}
as well as the integration region $B(\mathbf{r},V)$, which is a sphere of volume $V$ centered at the point $\mathbf{r}$. This point-detector is therefore equivalent to a single-mode detector with detector-mode $B(\mathbf{r},V)/\sqrt{V}$. Notably, if the region of integration was not small, then the corresponding number operator would not be single-mode. The Fock-projection operators which correspond to this measured mode are given by
\begin{equation}
    \hat{n}_{s,V}(N,M | \mathbf{r}) = :\left(\frac{\left(\hat{a}^\dagger_{s,V}(\mathbf{r})\right)^N\left(\hat{a}_{s,V}(\mathbf{r})\right)^M}{\sqrt{N!M!}}e^{-\hat{n}_{s,V}(\mathbf{r})}\right):,
\end{equation}
where here $:\cdot:$ represents normal-ordering, where all annihilation operators $\hat{a}_{s,V}(\mathbf{r})$ are placed on the right and all creation operators $\hat{a}^\dagger_{s,V}(\mathbf{r})$ are placed on the left. The point-detector is restricted to the observation of photons in this single-mode subsystem, interpreting the state at each point in terms of the expectation values of the various $\hat{n}_{s,V}(N,M | \mathbf{r})$. As with the single-mode case of the previous section, we will define a `photon-number subsystem' of the measured state $\hat{\rho}$ as
\begin{equation}
    \rho(N|\mathbf{r}) = \text{Tr}[\hat{\rho}\hat{n}_{s,V}(N,N | \mathbf{r})].
\end{equation}
This corresponds to the probability of measuring $N$ photons at position $\mathbf{r}$ with the point-detector. It now becomes apparent why we utilize the word `subsystem' when referring to the extracted probability of observing a particular multiphoton wavepacket, and it is because of this example where point-detectors are implemented. Each photon-number subsystem, as a function of $\mathbf{r}$, describes the probability of observing a given number of photons at each point in space. Furthermore, these subsystems together make up all necessary information for the measurement of number operators at each point in space. Now, recalling that the states $\hat{\rho}$ which we are interested in only contain photons of the mode $(s,f)$, we are able to express the state in terms of its Glauber $P$-function as
\begin{equation}
    \hat{\rho} = \int d^2\alpha P(\alpha)\ket{\alpha}\bra{\alpha}_{s,f},
\end{equation}
where $\ket{\alpha}_{s,f}$ is a coherent state of photons in the mode $(s,f)$. Noting the eigenvalue relation
\begin{equation}
    \hat{a}_{s,V}(\mathbf{r})\ket{\alpha}_{s,f} \approx \sqrt{V}\alpha f^*(\mathbf{r})\ket{\alpha}_{s,f},
\end{equation}
we can then conclude that
\begin{equation}
    \text{Tr}[\hat{\rho}\hat{n}_{s,V}(N,N | \mathbf{r})] \approx \int d^2\alpha P(\alpha)\frac{V^N |\alpha^* f(\mathbf{r})|^{2N}}{N!}e^{-V|\alpha f(\mathbf{r})|^2}.
\end{equation}
For a variety of states, this formula allows for the simplest computation of the state's subsystems as seen by a point-detector. The evolution of these distributions contains important information about the dynamics of photon-statistics. As an example, coherent states correspond to the Glauber $P$-function $\delta(\alpha_0 - \alpha)$ for some coherent amplitude $\alpha_0$. The resulting subsystems for such a state are then given by
\begin{equation}
    \rho^{\text{coh}}(N|\mathbf{r}) = \frac{V^N |\alpha_0^* f(\mathbf{r})|^{2N}}{N!}e^{-V|\alpha_0 f(\mathbf{r})|^2}.
\end{equation}
In this paper, we are able to measure all such subsystems of a coherent light field composed of Gaussian-beam photons. In doing so, we uncover a set of nontrivial quantum dynamics which give rise to an ultimately classical instance of the free electromagnetic field. Now, let us turn towards the notion of phase subsystems. Previously, we defined phase subsystems as the expectation values of operators of the form $\ket{\phi}\bra{\phi}/(2\pi)$. In the field-theoretic approach, concerning the phase subsystems observed by point-detectors, the corresponding operators are given by
\begin{equation}
    \hat{\Phi}_{s,V}(\phi|\mathbf{r}) = \sum_{n,m=0}^\infty e^{i\phi(n-m)}\hat{n}_{s,V}(n,m|\mathbf{r}).
\end{equation}
As such, we define phase subsystems in analogy to photon-number subsystems via
\begin{equation}
    \rho(\phi|\mathbf{r}) = \text{Tr}[\hat{\rho}\hat{\Phi}_{s,V}(\phi|\mathbf{r})].
\end{equation}
This quantity can be interpreted as the probability of observing the phase $\phi$ at position $\mathbf{r}$ with point-detector of polarization $s$ and volume $V$. In the case of a single-mode quantum state of mode-function $f$, we can once-again use its $P$-function representation to obtain
\begin{equation}
\begin{aligned}
    \rho(\phi|\mathbf{r}) &= \int d^2\alpha P(\alpha) \braket{\alpha|\hat{\Phi}_{s,V}(\phi|\mathbf{r})|\alpha}_{s,f}\\
    &= \int d^2\alpha P(\alpha) \sum_{n,m=0}^\infty e^{i\phi(n-m)}\braket{\alpha|\hat{n}_{s,V}(n,m|\mathbf{r})|\alpha}_{s,f}\\
    &\approx \int d^2\alpha P(\alpha) \sum_{n,m=0}^\infty e^{i\phi(n-m)}\frac{\sqrt{V}^{n+m}(\alpha^*f(\mathbf{r}))^n(\alpha f^*(\mathbf{r}))^m}{\sqrt{n!m!}}\exp(-V|\alpha^* f(\mathbf{r})|^2)\\
    &= \int d^2\alpha P(\alpha) \theta(\sqrt{V}\alpha^*f(\mathbf{r})e^{i\phi}),
\end{aligned}
\end{equation}
where here we have defined the function $\theta$, which maps complex numbers to real numbers, as
\begin{equation}
    \theta(x) = \sum_{n,m=0}^\infty \frac{(x)^n (x^*)^m}{\sqrt{n!m!}}e^{-|x|^2}.
\end{equation}
Unfortunately, this function does not have a simple analytical form. Furthermore, another representation of the state which makes the expectation value of the operator $\hat{\Phi}_{s,V}(\phi|\mathbf{r})$ behave more nicely is currently unknown. As such, we must rely on numerical integration techniques to compute $\rho_{s,V}(\phi|\mathbf{r})$. This function is also important for the case where the quantum state is coherent of complex amplitude $\alpha_0$, since this will result in
\begin{equation}
    \rho^{\text{coh}}(\phi|\mathbf{r}) = \theta(\sqrt{V}\alpha^*_0 f(\mathbf{r})e^{i\phi}).
\end{equation}
Luckily, $\theta(x)$ has a fairly simple distribution when evaluated numerically, allowing for analytical approximations to speed up numerical simulations. In particular, if $x = x_1+ix_2$, where $x_1,x_2$ are real, then we have
\begin{equation}
    \theta(x_1,x_2)\approx A(x_1)\exp\left[-\frac{x_2}{\sigma(x_1)}\right],
\end{equation}
where
\begin{equation}
    A(x_1) = \left(1-\frac{1}{c_A}\exp\left[-(x_1-d_A)^2\right]\right)\left(5x_1\left[\frac{1+\tanh(a_A(x_1-b_A))}{2}\right]+e^{2x_1}\left[\frac{1+\tanh(a_A(b_A-x_1))}{2}\right]\right)
\end{equation}
for parameters $a_A = 5.07, b_A = 0.418, c_A = 9.89,$ and $d_A = 1.12$. Additionally, we have defined
\begin{equation}
    \sigma(x_1) = \left(\frac{\tanh(-a_\sigma x_1)}{b_\sigma} + c_\sigma\right)\left(\frac{1+\tanh(A_\sigma x_1)}{2}\right) + \left(\frac{\tanh(-d_\sigma x_1)}{e_\sigma} + f_\sigma\right)\left(\frac{1+\tanh(-A_\sigma x_1)}{2}\right),
\end{equation}
where here we have $a_\sigma=0.466, b_\sigma=0.788,c_\sigma=1.78,d_\sigma=0.426,e_\sigma=0.754,f_\sigma=-0.358,$ and $A_\sigma=0.442$. With these analytical approximations to $\theta(x)$, it is numerically simple to evaluate the phase subsystems for a given quantum state.

\end{document}